\documentclass[fleqn,usenatbib]{mnras}

\usepackage{newtxtext,newtxmath}

\usepackage[T1]{fontenc}

\DeclareRobustCommand{\VAN}[3]{#2}
\let\VANthebibliography\thebibliography
\def\thebibliography{\DeclareRobustCommand{\VAN}[3]{##3}\VANthebibliography}

\usepackage{hyperref,enumerate}
\usepackage{graphicx}	
\usepackage{amsmath}	
\usepackage{multirow}
\usepackage{cuted}
\usepackage{longtable}
\usepackage{soul}
\usepackage[T1]{fontenc}
\usepackage{gensymb}
\usepackage{ulem}
\usepackage{verbatim}

\title[A transition from supercritical to thin disk]{The bright black hole X-ray binary 4U 1543–47 during 2021 outburst. A clear state transition from super-Eddington to sub-Eddington accretion revealed by \textit{Insight}-HXMT}

\author[P.\ Jin et al.]{
Pei Jin,$^{1,2}$
Guobao Zhang,$^{1,2}$
Yuexin Zhang,$^{3,4}$
Mariano M\'{e}ndez,$^{3}$
Jinlu Qu,$^{2,5}$
David M.\ Russell,$^{6}$
\newauthor{Jiancheng Wang,$^{1,2}$
Shuangnan Zhang,$^{2,5}$
Yi-Jung Yang,$^{7,8}$
Shumei Jia,$^{2,5}$
Zixu Yang,$^{9}$
and Hexin Liu$^{2,5}$
}
\\
$^{1}$Yunnan Observatories, Chinese Academy of Sciences, Kunming 650216, People's Republic of China\\
$^{2}$University of Chinese Academy of Sciences, Beijing 100049, People's Republic of China\\
$^{3}$Kapteyn Astronomical Institute, University of Groningen, P.O.\ BOX 800, 9700 AV Groningen, The Netherlands\\
$^{4}$Center for Astrophysics, Harvard \& Smithsonian, 60 Garden St, Cambridge, MA 02138, USA\\
$^{5}$Key Laboratory of Particle Astrophysics, Institute of High Energy Physics, Chinese Academy of Sciences, Beijing 100049, China\\
$^{6}$Center for Astrophysics and Space Science (CASS), New York University Abu Dhabi, PO Box 129188, Abu Dhabi, UAE\\
$^{7}$Department of Physics, The University of Hong Kong, Pokfulam Road, Hong Kong\\
$^{8}$Laboratory for Space Research, The University of Hong Kong, Hong Kong\\
$^{9}$School of Physics and Optoelectronic Engineering, Shandong University of Technology, Zibo 255000, China\\
}

\date{Accepted 2024 March 4. Received 2024 March 1; in original form 2023 February 8}

\pubyear{2024}

\begin{document}
\label{firstpage}
\pagerange{\pageref{firstpage}--\pageref{lastpage}}
\maketitle

\begin{abstract}
We present a detailed analysis of the observations with the Hard X-ray Modulation Telescope of the black hole X-ray transient 4U~1543$-$47 during its outburst in 2021.
We find a clear state transition during the outburst decay of the source. 
Using previous measurements of the black-hole mass and distance to the source, the source luminosity during this transition is close to the Eddington limit. 
The light curves before and after the transition can be fitted by two exponential functions with short ($\sim 16$ days) and long ($\sim 130$ days) decay time scales, respectively.
We detect strong reflection features in all observations that can be described with either the \textsc{relxillns} or \textsc{reflionx\_bb} reflection models, both of which have a black-body incident spectrum. 
In the super-Eddington state, we observe a Comptonized component characterized by a low electron temperature of approximately 2.0 keV. We suggest that this component appears exclusively within the inner radiation-pressure dominated region of the supercritical disk as a part of the intrinsic spectrum of the accretion disk itself. This feature vanishes as the source transitions into the sub-Eddington state.
The emissivity index of the accretion disk in the reflection component is significantly different before and after the transition, $\sim3.0-5.0$ and $\sim7.0-9.0$ in the super- and sub-Eddington states, respectively.
Based on the reflection geometry of returning disk radiation, the geometrically thicker the accretion disk, the smaller the emissivity index.
Therefore, we propose that the transition is primarily driven by the change of the accretion flow from a supercritical to a thin disk configuration.

\end{abstract}




\section{INTRODUCTION}
\label{sec:introduction}

Black hole X-ray binaries (BHXBs) are typically transient systems that stay in the quiescent state most of the time, but show recurrent bright X-ray outbursts lasting for weeks to months~\citep[e.g.][]{1996ARA&A..34..607T, 2006csxs.book..157M}. It is generally believed that the outbursts are due to the thermal instability of the accretion disk~\citep[e.g.][]{2001NewAR..45..449L}.

During an outburst, the luminosity of a source usually rises rapidly and then decreases slowly~\citep[e.g.][]{1996ARA&A..34..607T}. 
In the hardness-intensity diagram (HID) the source traces a ``q'' shape during the outburst~\citep[e.g.][]{2004MNRAS.355.1105F,2005Ap&SS.300..107H,2011BASI...39..409B}, evolving from the low-hard state (LHS), the hard-intermediate state (HIMS), the soft-intermediate state (SIMS) and the high-soft state~\citep[HSS; e.g.][]{1997ApJ...479..926M,2005A&A...440..207B}.
The time-averaged energy spectrum can be described by the combination of a thermal and a power-law (PL) component~\citep[e.g.][]{2006csxs.book..157M}. 
In the LHS, the spectrum is dominated by the PL component, which could be produced by inverse-Compton scattering of soft photons from the accretion disk~\citep{1973A&A....24..337S} in a hot corona with temperature up to a few 100~keV~\citep[e.g.][]{1980A&A....86..121S}. The spectrum in the HSS is dominated by the emission from the accretion disk, which could be described by multi-temperature blackbody~\citep[][]{1984PASJ...36..741M}.

A black hole transient generally shows transitions from the LHS to the HSS around the peak of the outburst~\citep[e.g.][]{2011BASI...39..409B}.
However, sometimes new branches appear in the HID around the peak of the outburst when the outburst is very bright, often reaching luminosities above the Eddington limit.
Examples of this are the ``anomalous'' state~\citep{2010LNP...794...53B} and the ``steep powerlaw'' state in the 1996 outburst of GRO~J1655$-$40~\citep{2006ARA&A..44...49R}, or the ``ultra-luminous'' state~\citep{2012MNRAS.427..595M} and the ``hypersoft'' state~\citep{2015MNRAS.451..475U} in the 2005 outburst of GRO~J1655$-$40.
A supercritical accretion disk is expected when the emission exceeds the Eddington limit~\citep[e.g.][]{1973A&A....24..337S, 2007MNRAS.377.1187P, 2017ARA&A..55..303K, 2021AstBu..76....6F}.
Due to the radiation pressure in the supercritical accretion disk model, the inner regions of the disk become geometrically thick, $H/R \sim 1$, where $R$ is the distance to the black hole and $H$ is the half-thickness of the disk~\citep{2007MNRAS.377.1187P, 2021AstBu..76....6F}. 
An outflow in the form of a powerful wind is also expected in the supercritical accretion disk~\citep{1973A&A....24..337S, 2007MNRAS.377.1187P, 2021AstBu..76....6F}.
The supercritical accretion disk model has been successfully used to explain observations of Ultraluminous X-Ray Sources~\citep[ULXs;][]{2017ARA&A..55..303K, 2021AstBu..76....6F}. After considering the effect of advection or mass loss, the disk becomes slim, $H/R\leq 1$, rather than geometrically thick~\citep{1988ApJ...332..646A, 1999AstL...25..508L, 2016A&A...587A..13L, 2017ARA&A..55..303K}.

A common observation in ULXs is the presence of a soft excess, a phenomenon that can be effectively characterized by Compton upscattering involving a low electron temperature, typically $\sim$2.0 keV~\citep{2009MNRAS.397.1836G, 2021AstBu..76....6F}.
The soft excess maybe be an intrinsic part of the spectrum of the accretion disc itself~\citep{2004ApJ...601..405S, 2012MNRAS.420.1848D}, possibly caused by the Magnetorotational Instability (MRI) in the inner radiation-pressure dominated region of the accretion disk when the mass accretion rate approaches the Eddington limit~\citep{2004ApJ...601..405S}, which drives large-scale turbulence.
The MRI creates conditions that lead to Comptonization, potentially enabling this process to take place within the disk itself~\citep{2004ApJ...601..405S, 2012MNRAS.420.1848D}.  
If the soft excess is indeed from the accretion disc itself, it is likely that it will be powered by the mass accretion~\citep{2012MNRAS.420.1848D}. 
Consequently, the presence of the soft excess is expected to become more pronounced as the accretion rate approaches the Eddington limit~\citep{2012MNRAS.420.1848D}.

A broad asymmetric iron fluorescence line that is sometimes accompanied by a hump at 20--30 keV is commonly observed in the hard state of BHBs~\citep{2014SSRv..183..277R, 2021SSRv..217...65B}, such as Cyg X$-$1~\citep{2017MNRAS.472.4220B} and MAXI J1535$-$571~\citep{2018ApJ...852L..34X}. This feature is interpreted as photons from the corona being reflected off the accretion disk~\citep{2021SSRv..217...65B}. When reflection occurs in the region near the black hole, the profile of the iron fluorescence line is broadened by gravitational and Doppler effects~\citep{2016A&A...590A..76D}.

In recent years, reflection in the soft state in some BHBs has also been reported~\citep{2020ApJ...892...47C, 2021ApJ...909..146C, 2021ApJ...906...11W, 2021ApJ...921..155L}; because in this state the spectrum is dominated by the emission of the accretion disk, this reflection cannot be explained by irradiation of the disk from the PL component. A model, \textsc{relxillns}~\citep{2022ApJ...926...13G, 2022MNRAS.514.3965D}, was developed to fit the reflection in accretion disks around neutron stars. 
The irradiating spectrum in the \textsc{relxillns} model is a single-temperature blackbody.
\cite{2020ApJ...892...47C} found that the reflection spectrum in a very soft state of the BHB XTE J1550$-$564 can be well described by the \textsc{relxillns} model. 
The \textsc{relxillns} model has also been successfully used to describe the reflection in the soft state in the BHBs 4U 1630$-$47 and MAXI J0637$-$430~\citep{2021ApJ...909..146C, 2021ApJ...921..155L}.
The reflection feature in the soft state is interpreted as being produced by disk photons that bend back by the strong gravity of the black hole and illuminate the surface of the accretion disk when the inner radius of the disk reaches the innermost stable circular orbit (ISCO).

4U~1543$-$47 is a transient galactic source with a dynamically-confirmed BH primary~\citep{2006ARA&A..44...49R}. The black hole mass is 9.4 $\pm$ 1.0 $M_\odot$ and the mass of the companion star is 2.45 $\pm$ 0.15 $M_\odot$; the distance to the source is 7.5 $\pm$ 0.5 kpc~\citep{2002AAS...201.1511O, 2004MNRAS.354..355J, 2006MNRAS.371.1334R}. 
The orbital period of the system is 26.8 hrs and the orbital inclination angle is $20.7^{\circ} \pm 1.5^{\circ}$~\citep{2003IAUS..212..365O}. An earlier measurement derived an orbital inclination angle of $30^{\circ} \pm 6^{\circ}$~\citep{1998ApJ...499..375O}. The accretion disk inclination angle obtained from fits to the X-ray spectra ranges from $\sim 32^{\circ}$~\citep{2014ApJ...793L..33M} to $\sim 36^{\circ}$~\citep{2020MNRAS.493.4409D}.

The Monitor of All-sky X-ray Image (MAXI)~\citep{2009PASJ...61..999M} Gas Slit Camera (GSC) detected a new outburst of 4U~1543$-$47 on June 11, 2021~\citep{2021ATel14701....1N}; before that, the source had stayed in the quiescent state for $\sim 20$ years. Since it was first discovered in 1971~\citep{1972ApJ...174L..53M}, this is the fifth outburst of 4U~1543$-$47; the other four outbursts were in 1971, 1983, 1992, 2002~\citep{2004ApJ...610..378P}. 
On the same day when it was detected, the source entered the soft state~\citep{2021ATel14701....1N} until January 3, 2022, when a radio counterpart was detected~\citep{2022ATel15157....1Z} by the South African MeerKAT radio telescope, suggesting that the source entered the hard state. 
The peak flux of the \textit{MAXI}/GSC 2--10 keV light curve reached about 8.2 Crab~\citep{2021ATel14708....1N}, suggesting that the source reached the Eddington luminosity. 
Using observations with the Neutron star Interior Composition Explorer (NICER) an iron emission line at $\sim$ 6.4 keV was detected in the soft state~\citep{2021ATel14725....1C}.

In this paper, we study the spectral evolution of the BHB 4U~1543$-$47 during the new outburst in 2021 with the Hard X-ray Modulation Telescope~\citep[\textit{Insight}-HXMT;][]{2020SCPMA..6349502Z}. The paper is organized as follows: In section \ref{sec:OBSERVATION AND DATA REDUCTION} we describe the data reduction and analysis with \textit{Insight}-HXMT. In section \ref{sec:RESULTS} we show the results from light curves and spectra, and in section \ref{sec:discussion} we discuss our results.

\section{OBSERVATION AND DATA REDUCTION}
\label{sec:OBSERVATION AND DATA REDUCTION}

Launched on 2017 June 15, \textit{Insight}-HXMT~\citep{2020SCPMA..6349502Z} is the first Chinese X-ray observatory. \textit{Insight}-HXMT carries three telescopes: the Low Energy (LE) telescope (1--15 keV), the Medium Energy (ME) telescope (5--30 keV), and the High Energy (HE) telescope (20--250 keV). Starting on June 14, 2021, four days after MAXI detected the outburst of 4U~1543$-$47, \textit{Insight}-HXMT made a total of 54 observations and produced 135 sub-exposures.
We used the \textit{Insight}-HXMT data analysis software \textit{Insight}-HXMTDAS (version 2.05) and the latest CALDB files (version 2.06) to process the \textit{Insight}-HXMT data. We used only data from the small-FOV detectors. We created good time intervals with the following selection criteria: (1) pointing offset angle $< 0.04^{\circ}$, (2) elevation angle $> 10^{\circ}$, (3) geomagnetic cutoff rigidity $> 8$~GV, and (4) the satellite was at least $300$~s away from the crossing of the South Atlantic Anomaly (SAA). We created background spectra with the {\sc lebkgmap, mebkgmap}, and {\sc hebkgmap} tasks in \textit{Insight}-HXMTDAS, and response files with the {\sc lerspgen, merspgen}, and {\sc herspgen} tasks, respectively. 
In the data reduction of \textit{Insight}-HXMT/ME, the number of pixels operated to generate science data in each sub-exposure is different, especially during high count-rate periods, resulting in different effective areas during the outburst. To solve this problem, we use the pixel-average photon count rate to generate the 10--30 keV light curve.

Discarding the data with exposure times less than 200s in LE and ME, we used 121 sub-exposures in total for the spectral analysis.
We extracted spectra from LE, ME, and HE separately in every single sub-exposure. 
Due to the low signal-to-noise, the HE data with exposure times less than 200s in the following nine sub-exposures are not used in the spectral analysis: P030402600102, P030402600107, P030402600116, P030402600123, P030402600213, P030402600302, P030402600307, P030402601203 and P030415501301. 
We combined the LE (2--10 keV), ME (10--30 keV), and HE (28--60 keV) spectra in each sub-exposure.
We added 1\%, 2\%, and 1\% systematic errors to the LE, ME, and HE spectra, respectively. 
Using \textsc{grppha}, we rebinned the spectra as follows: For LE, from 2 keV to 7 keV the data are grouped to 4 bins in each group, while from 7 keV to 10 keV the data are grouped to 16 bins in each group. For ME, from 10 keV to 20 keV the data are grouped to 4 bins in each group, while from 20 keV to 30 keV the data are grouped to 16 bins in each group. For HE, we group the data to 4 bins in each group.
The software package {\sc XSPEC}~\citep{1996ASPC..101...17A}, version 12.12.0, is used in our spectral analysis. 
We use the model \textsc{tbabs} to describe the absorption due to the interstellar medium along the line of sight, with the solar abundance tables of~\citet{2000ApJ...542..914W} and the cross-section tables of~\citet{1996ApJ...465..487V}.
We use a multiplicative scaling factor (\textsc{constant} in \textsc{XSPEC}) to account for slightly different normalizations of LE, ME, and HE. 
All errors represent the 68\% confidence range for a single parameter unless otherwise stated.

We use the Python software library \textsc{stingray} \citep{2019ApJ...881...39H} to generate power density spectra (PDS) in the 1--10 keV (LE), 10--30 keV (ME) and 28--60 keV (HE) bands. We generate averaged PDSs using 128 s duration segments from the sub-exposure light curves with a time resolution of 0.01 s and bin up in frequency over a minimum geometric spacing of 1.05. The PDSs show low-level variability at low frequencies (0.01$-$10.0 Hz) and have no visible QPOs in the 1$-$60 keV. We will present a detailed timing analysis in Jin et al. (in preparation).

\section{DATA ANALYSIS AND RESULTS}
\label{sec:RESULTS}

\subsection{Light Curve and HID}
\label{sec:Light Curve and HID}

\begin{figure*}
    \vspace{-15mm}
	\includegraphics[width=\textwidth]{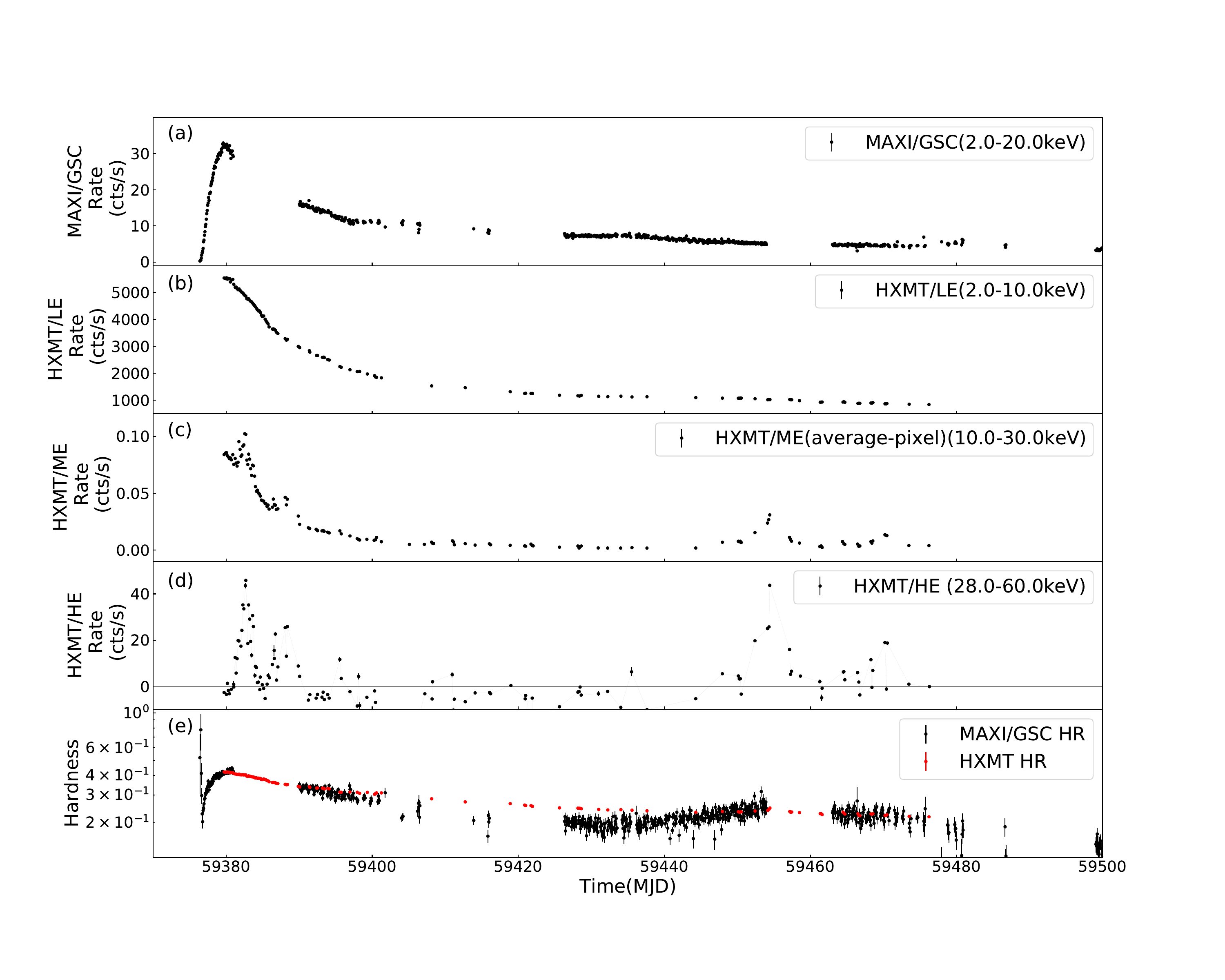}
	\vspace{-15mm}
    \caption{ Light curves and hardness curves of 4U 1543$-$47 (a) MAXI/GSC 2--20 keV light curve. (b) \textit{Insight}-HXMT/LE 2--10 keV light curve. (c) \textit{Insight}-HXMT/ME 10--30 keV light curve. (d) \textit{Insight}-HXMT/HE 28--60 keV light curve. (e) Black dots, hardness ratio of MAXI/GSC (4--10 keV)/(2--4 keV), and red dots, hardness ratio of \textit{Insight}-HXMT/LE (4--10 keV)/(2--4 keV).}  \vspace{-2mm}
    \label{fig:lc}
\end{figure*}

The 2--20 keV 1-orbit average \textit{MAXI}/GSC light curve of the outburst of 4U~1543$-$47 (Figure~\ref{fig:lc}, panel (a)) shows a typical fast rise and a gradual decay. The outburst started on June 11 2021 (MJD 59376), reached a maximum of $\sim 33$ counts/s on June 14 2021 (MJD 59379), and decayed exponentially.
To compare the peak intensity of this outburst with that of the earlier outbursts of 4U~1543$-$47 (see section \ref{sec:Comparison with Earlier Outbursts}), we obtained the 3.7--7.5 keV MAXI/GSC peak intensity\footnote{http://maxi.riken.jp/mxondem/}: $\sim$ 12 counts/s, about 10 Crab.
The 2--10 keV \textit{Insight}-HXMT/LE light curve is shown in panel (b) where each point represents one sub-exposure. The LE light curve missed the rising phase of the outburst, and the maximum intensity at the beginning is $\sim 5600$ counts/s. Compared with the MAXI light curve, the first \textit{Insight}-HXMT observation was performed around the peak of the outburst.
The \textit{Insight}-HXMT/ME 10--30~keV pixel-averaged light curve is shown in panel (c) of Figure~\ref{fig:lc}. Unlike the smooth light curve in the 2--10 keV band, the pixel-averaged ME light curve shows some hard flares on top of the whole outburst.
The \textit{Insight}-HXMT/HE 28--60~keV light curve is dominated by several hard flares (Figure~\ref{fig:lc}, panel (d)). The times of the hard flares in the HE light curve are consistent with those in the ME light curve. The HE observation with a maximum intensity of $\sim 50$ counts/s happens at the peak of the first hard flare and appears after the peaking time of the outburst. The HE observations away from the hard flares are dominated by the background.
To investigate the evolution of the energy spectrum, we calculate the hardness ratio (HR) as the count-rate ratio in the 4--10 keV and 2--4 keV energy bands. The HR calculated from MAXI and \textit{Insight}-HXMT/LE is shown in panel (e) of Figure~\ref{fig:lc} with black and red-filled circles, respectively. The HR from MAXI and \textit{Insight}-HXMT/LE, are consistent during the outburst.
In the rising phase of the outburst, the HR of MAXI decreases very rapidly from $\sim 0.8$ down to $\sim 0.2$, indicating that the source enters the soft state immediately after the outburst starts. The fast hard-to-soft state transition also appeared at the beginning of the outburst in 2002 in 4U~1543$-47$~\citep{2004ApJ...610..378P}. 
After the minimum point, the HR increases with intensity and decreases slowly after reaching the peak ($HR \sim 0.4$).

\begin{figure}
    \vspace{-5mm}
	\includegraphics[width=\columnwidth]{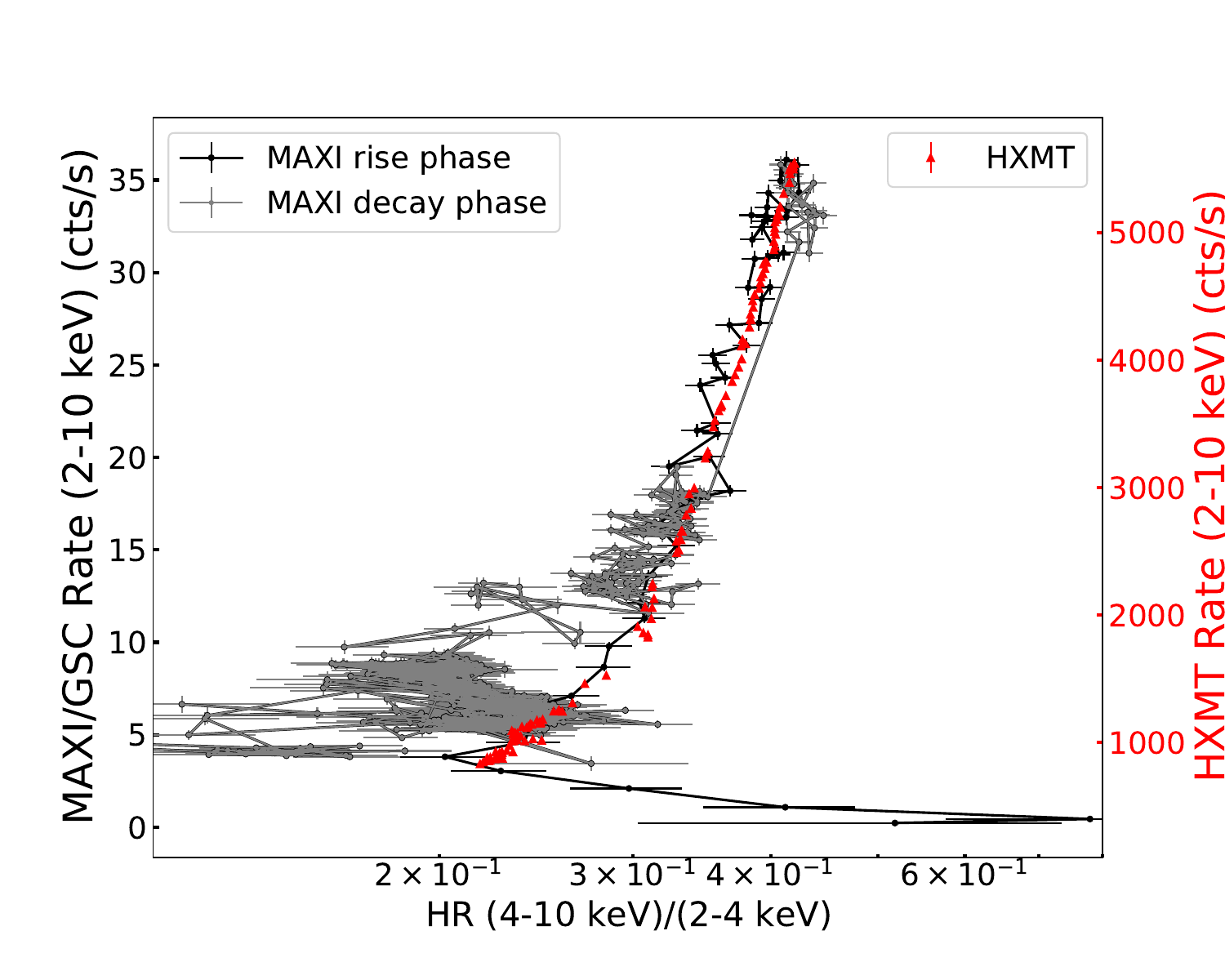}
	\vspace{-5mm}
    \caption{Hardness intensity diagram of 4U~1543$-$47 observed with \textit{MAXI}/GSC (black/gray circles) and \textit{Insight}-HXMT/LE (red triangles).}  
    \label{fig:HID}
\end{figure}

In Figure~\ref{fig:HID} we plot the hardness-intensity diagrams (HID) for the 2021 outburst of 4U~1543$-$47 observed by MAXI (black/gray circles) and \textit{Insight}-HXMT/LE (red triangles). Unlike hardness-intensity tracks commonly observed in other BHBs systems where the source transitions from the LHS to the HSS at the peak of the outburst~\citep{2006ARA&A..44...49R, 2004MNRAS.355.1105F, 2011BASI...39..409B}, the shape of the MAXI HID of 4U~1543$-47$ is odd near the peak, absent of horizontal evolution at the peak.
After the outburst starts, the source moves from the lower right to the lower left and then moves to the upper middle on the HID. During the outburst decay, the source moves from the upper middle to the lower left. 
The shape of HID near the peak is similar to the ``ultra-luminous'' state of GRO~J1655$-$40 in the 2005 outburst~\citep{2012MNRAS.427..595M}, a ``diagonal bar'' shape near maximum intensity.
The \textit{Insight}-HXMT/LE HID also shows the ``diagonal bar'' shape near the peak.
The complete HID that shows a "q" loop including the hard state, using MAXI and NICER data, will be presented in Alnaqbi et al. (in preparation).

\subsection{Spectral Analysis}
\label{sec:Spectral Analysis}

\subsubsection{Initial  Model selection}
\label{sec:Model selection}

\begin{figure}
    \vspace{-5mm}
	\includegraphics[width=\columnwidth]{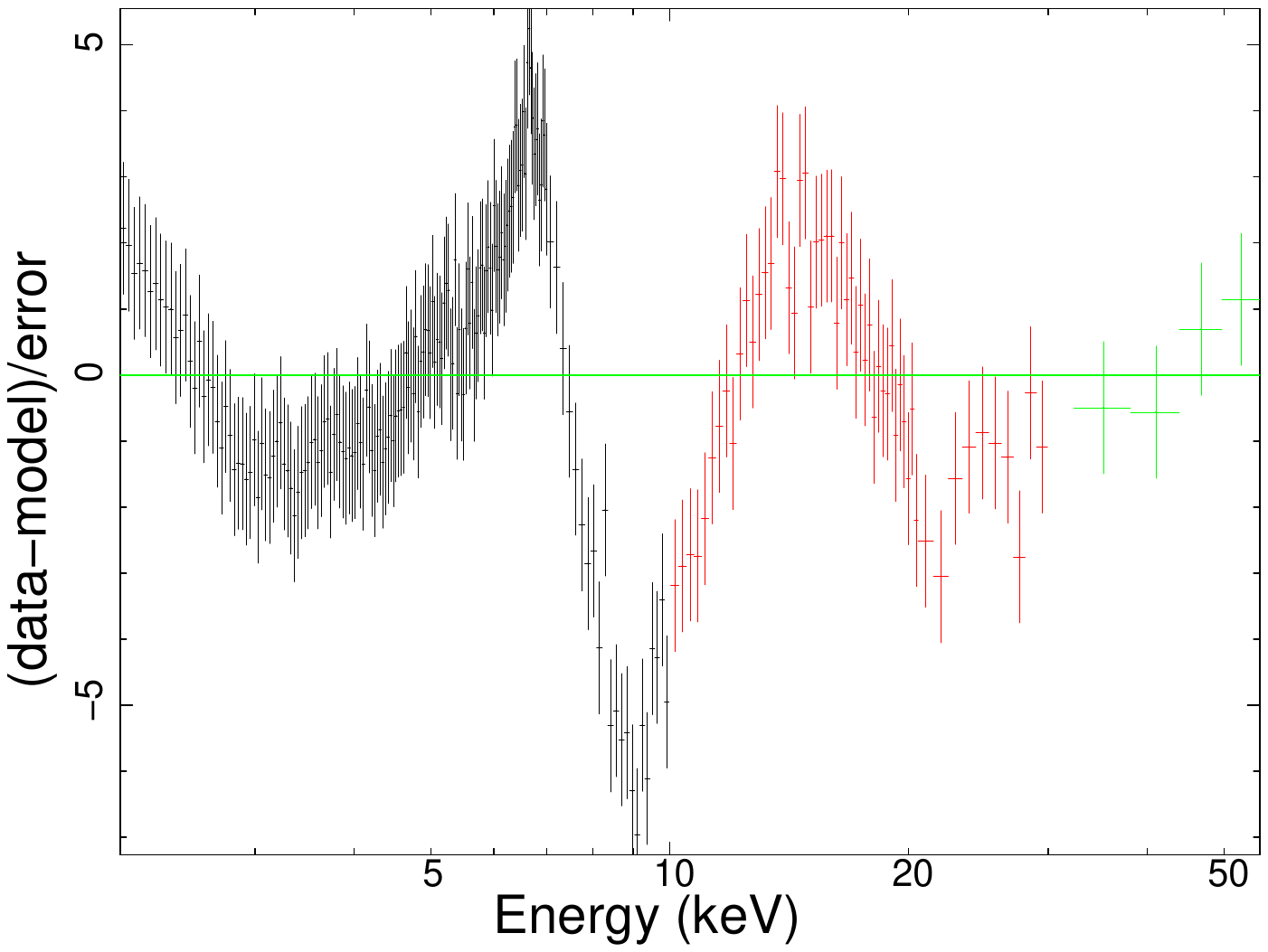}
    \vspace{-5mm}
    \caption{Fitting residuals of the 2--60 keV \textit{Insight}-HXMT/(LE$+$ME$+$HE) spectra of the
    sub-exposure P030402600120 (MJD 59382) of 4U~1543$-$47. The model used is \textsc{constant*tbabs(thcomp*diskbb)}.}  
    \label{fig:fig3}\vspace{-2mm}
\end{figure}

To find the proper components for the energy spectra, we take \textit{Insight}-HXMT sub-exposure P030402600120 for the initial spectral fitting. This observation was carried out on MJD 59382, near the peak of the outburst, and from all our observations this is one of the spectra with the best statistics.
We first tried the phenomenological model \textsc{constant*tbabs(thcomp*diskbb)} to fit the spectra. The \textsc{diskbb} component describes the thermal emission from the accretion disk. We used the convolution component \textsc{thcomp}~\citep{2020MNRAS.492.5234Z} to account for the non-thermal emission from the corona. 
A scaling factor is introduced to account for the different normalizations between LE/ME/HE instruments. This parameter is set to unity for LE and left free for ME/HE.
The best-fit residuals are shown in Figure~\ref{fig:fig3}, and the $\chi^2$ is 1002 for 221 degrees of freedom (d.o.f.) (2436/221 without the LE/ME systematic errors).
In the figure it is apparent that there is an asymmetric broad iron line at 6--7 keV, accompanied by a hump at 10--20 keV and a soft excess below 2.5 keV, implying the existence of a reflection component~\citep[e.g.][]{2021SSRv..217...65B}.

\begin{figure}
	\includegraphics[width=\columnwidth]{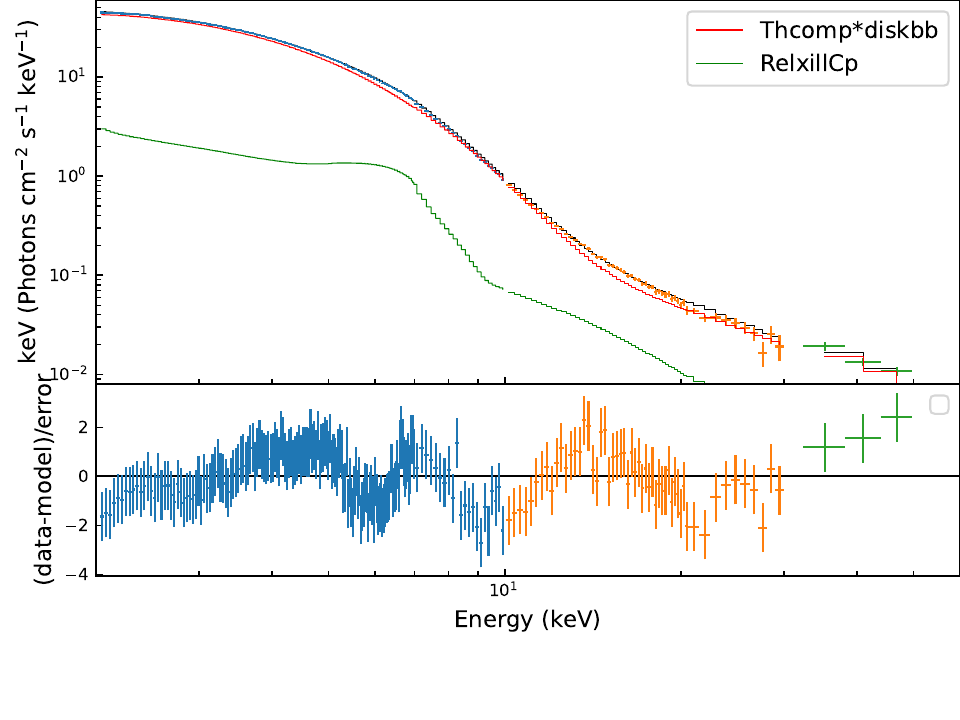}\vspace{-5mm}
 	\includegraphics[width=\columnwidth]{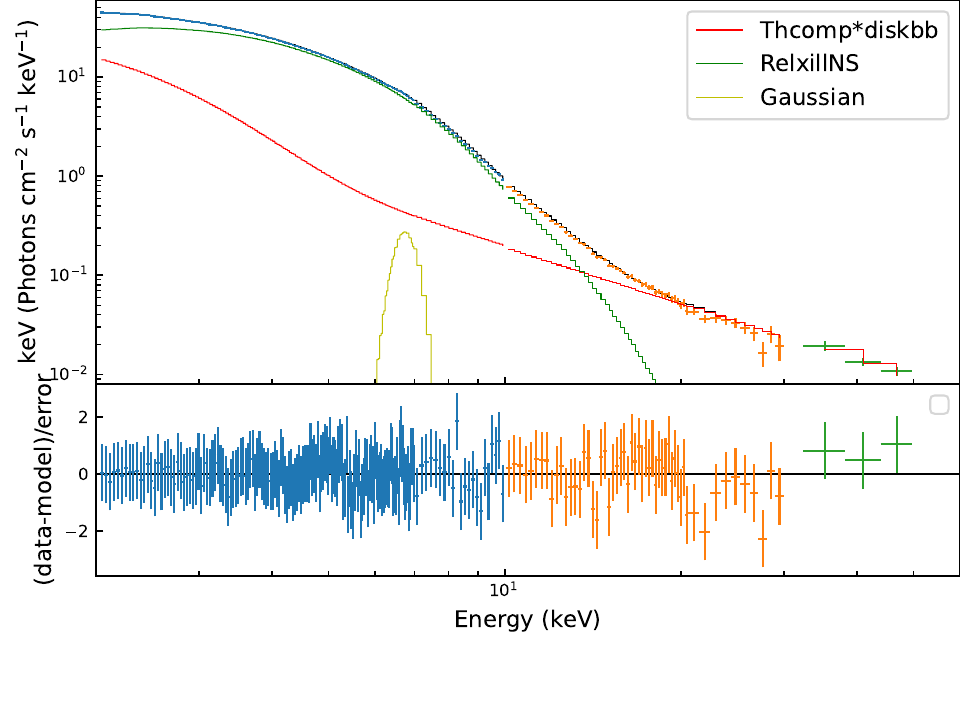}
	\vspace{-10mm}
    \caption{The unfolded spectrum and the residuals of sub-exposure P030402600120 (MJD 59382) of 4U~1543$-$47. 
    Top panel: the model is \textsc{constant*tbabs*(thcomp*diskbb+relxillcp)}.  
    Bottom panel: the model is \textsc{constant*tbabs*(thcomp*diskbb+relxillns+gaussian)}. The contributions of the various additive components to the model are also plotted.}
    \label{fig:fig4}\vspace{-2mm}
\end{figure}

To improve the fitting, we first added a \textsc{relxillcp} component~\citep{2014MNRAS.444L.100D, 2014ApJ...782...76G} to account for the irradiation of the accretion disk by a Comptonization continuum\footnote{We link the parameters $\Gamma$ and $kT_{e}$ of \textsc{relxillcp} to that of \textsc{thcomp} and fix $kT_{e}$ to the default (60 keV).}, and the $\chi^2/{\rm d.o.f.}$ decreases to 236/215 (879/215 without the LE/ME systematic errors) (Figure~\ref{fig:fig4}, top panel). The residuals, however, look wavy, with several structures.
The low $\chi^2/{\rm d.o.f.}$ implies that the systematic errors are overestimated.
The spectra of all observations are too soft to have enough high energy photons to illuminate the accretion disk and produce strong reflection in the case of the model \textsc{relxillcp}. This is particularly evident in observations outside of the hard flares shown in the HE and ME light curves, where photon counts above 30 keV are lacking.

Recently, reflection in the soft state in some BHBs has been reported~\citep{2020ApJ...892...47C, 2021ApJ...909..146C, 2021ApJ...906...11W, 2021ApJ...921..155L}. 
The reflection in the soft-state spectra in these BHBs is well fitted with the \textsc{relxillns} model. 
In the \textsc{relxillns} model, the reflection can be interpreted as being produced by returning disk radiation.
Since the spectra of 4U~1543$-$47 are very soft in all observations, we then used \textsc{constant*tbabs*(thcomp*diskbb+relxillns)} to fit the X-ray spectra; the $\chi^2/{\rm d.o.f.}$ is 105/215 (344/215 without the LE/ME systematic errors). 
\cite{2023arXiv231209632Z} analyzed five simultaneous observations of 4U~1543$-$47 with NuSTAR and Swift/XRT during the outburst decay. Also in their case, the fits to the broad-band (0.5$-$70 keV) X-ray spectra show that the model \textsc{relxillns} works better than the \textsc{relxillcp} at different flux levels.
There are still some positive residuals at energies between 6 and 7 keV. We then add a \textsc{gaussian} and get a $\chi^2/{\rm d.o.f.}$ of 81/213 (268/213 without the LE/ME systematic errors). 
The small $\chi^2/{\rm d.o.f.}$ is caused by the large systematic errors; see Applendix \ref{sec:systematic errors} for details. The best-fitting model and residuals are shown in the bottom panel of Figure~\ref{fig:fig4}.

\begin{figure}
	\includegraphics[width=\columnwidth]{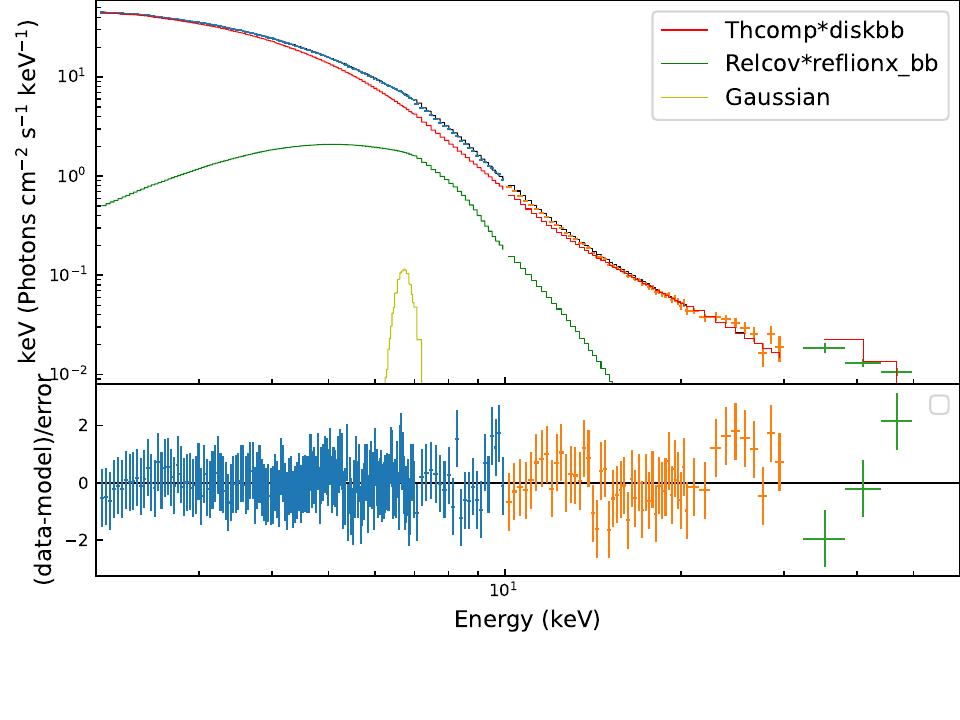}\vspace{-5mm}
 	\includegraphics[width=\columnwidth]{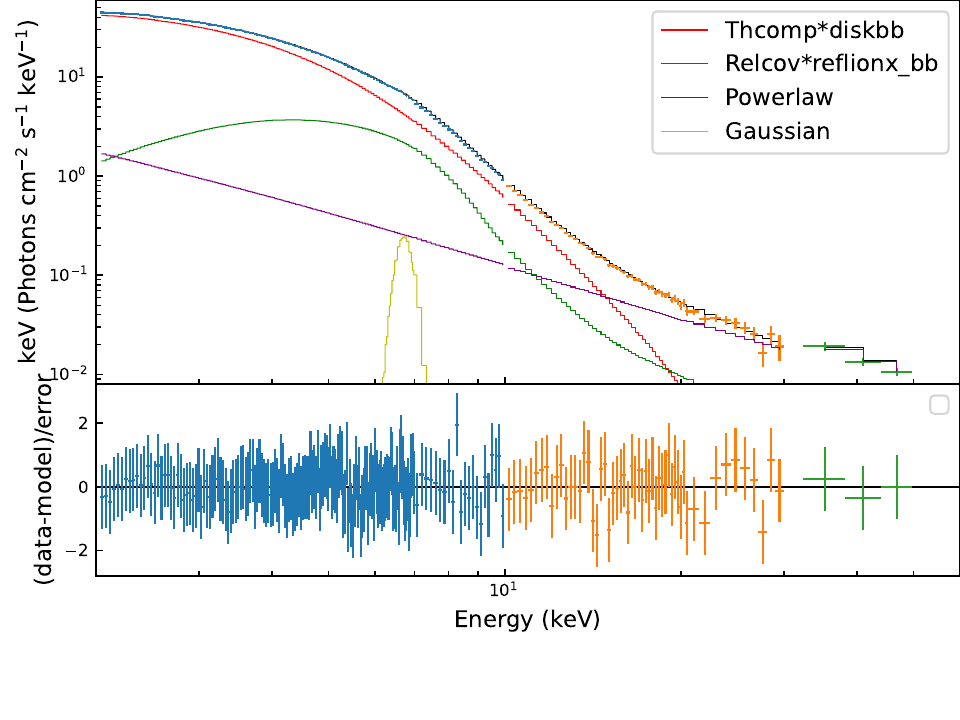}
	\vspace{-10mm}
    \caption{The unfolded spectrum and the residuals of sub-exposure P030402600120 (MJD 59382) of 4U~1543$-$47. 
    Top panel: the model is \textsc{constant*tbabs*zxipcf*(thcomp*diskbb+relconv*reflionx\_bb+gaussian)}.  
    Bottom panel: the model is \textsc{constant*tbabs*zxipcf*(thcomp*diskbb+} \textsc{powerlaw+relconv*reflionx\_bb+gaussian)}. The contributions of the various additive components to the model are also plotted.}\vspace{-2mm}
    \label{fig:fig4_reflionx}
\end{figure}

We use the \textsc{cflux} convolution model to caculate the fluxes in \textsc{xspec}.
The total flux in the 0.1--60 keV range is $\rm 3.87\times10^{-7} ergs\ cm^{-2}\ s^{-1}$, while the reflection flux within the same energy range is $\rm 2.50\times10^{-7} ergs\ cm^{-2}\ s^{-1}$. It is noteworthy that the reflection component contributes over 60\% of the total flux in the 0.1--60 keV band, or approximately 46\% when considering the interval from 0.1--2.0 keV.
This prompts us to consider whether the \textsc{relxillns} model might be overestimating  reflection photons, some of which might actually originate from the accretion disk. 
Subsequently, we explored the alternative reflection model \textsc{reflionx\_bb}~\citep{2017ApJ...836..140L} to fit the spectra. Similar to \textsc{relxillns}, the \textsc{reflionx\_bb} model is an X-ray reflection model using a black body as the input spectrum. 
With \textsc{reflionx\_bb} an absorption model \textsc{zxipcf}~\citep{2006A&A...453L..13M, 2008MNRAS.385L.108R} is required to improve the fit at low energies.
Consequently, the model is \textsc{constant*tbabs*zxipcf*(thcomp*diskbb+relconv*reflionx\_bb+} \textsc{gaussian)}.
In the top panel of Figure~\ref{fig:fig4_reflionx}, we show the result of the fit of the data of sub-exposure P030402600120, with a $\chi^2/{\rm d.o.f.}$ of 98/215 (314/215 without the LE/ME systematic errors).

\textsc{zxipcf} is used to describe the partial absorption by ionized material.
\citet{2023MNRAS.524.5817H} also employed \textsc{zxipcf} to improve their fits of the AstroSat spectra of the source, even thought the reflection model they used is \textsc{relxill}.
While with \textsc{relxillns} an additional absorber is not needed, this difference comes from whether the reflected emission is important at low energies.
Using \textsc{relxillns} the reflected emssion is prominent at low energies, whereas with \textsc{reflionx\_bb} or \textsc{relxill}, the contribution of the reflected emission is negligible at low energies (see Figure~\ref{fig:fig4}, ~\ref{fig:fig4_reflionx}, and section~\ref{sec:The super-Eddington state}).

Even though a low $\chi^2/{\rm d.o.f.}$ has been obtained, we notice that the fit is not very good above 20 keV, and we suggest that the low $\chi^2/{\rm d.o.f.}$ is caused by the large systematic errors, not overfitting.
Adding a \textsc{powerlaw}, the $\chi^2/{\rm d.o.f.}$ decreases to 66/214 (264/214 without the LE/ME systematic errors) (Figure~\ref{fig:fig4_reflionx}, bottom panel), where a low electron temperature of \textsc{thcomp}, $\sim$2.0 keV, is obtained.
The reflection contributes less than 10\% of the flux in the 0.1--60 keV band.

We test every component for all observations. We find that not all observations require two Comptonized components. For observations outside of the hard flares at the early decay of the outburst, \textsc{thcomp} with kT$_{\rm e} \sim$2.0 keV is enough. We will call this Comptonized component with low electron temperature the "soft excess".

In the following analysis, we employ the two aforementioned reflection models to separately fit all the data and investigate the evolution of parameters. 
We show the complete models in Table~\ref{tab:model}.

\onecolumn
\begin{strip}
\vspace{-10mm}
\begin{longtable}{p{12cm}}
    \caption{Models used to fit all the data.}
    \label{tab:model} \\
\hline 
\textrm{I} = \textsc{constant*tbabs*(thcomp*diskbb+relxillns+gaussian)}\\
\\
\textrm{II} = \textsc{constant*tbabs*zxipcf*(thcomp*diskbb+powerlaw+relconv*reflionx\_bb+gaussian)} \\
\hline
\textbf{Notes}: In model \textrm{I} we only use the \textsc{thcomp} component to fit all the hard photons.
As a result, the evolution of the parameters of \textsc{thcomp} are discontinuous (Figure~\ref{fig:fig5_relxillns}). \\
In model \textsc{II} the \textsc{thcomp} component describes the soft excess, while the \textsc{powerlaw} component describes the hard tail. The soft excess appears above the Eddington limit and disappears under the Eddington limit (Figure~\ref{fig:thcomp_frac}). The \textsc{powerlaw} component traces the HE 28.0--60.0 keV light curve (Figure~\ref{fig:fig5_reflionx}). Therefore, this suggests that they have completely different physical origins.
\end{longtable}
\end{strip}
\twocolumn

\subsubsection{The evolution of parameters with model \textrm{I}}
\label{sec:The evolution of parameters with relxillns}

\begin{figure*}
    \vspace{-40mm}
	\includegraphics[width=0.9\textwidth]{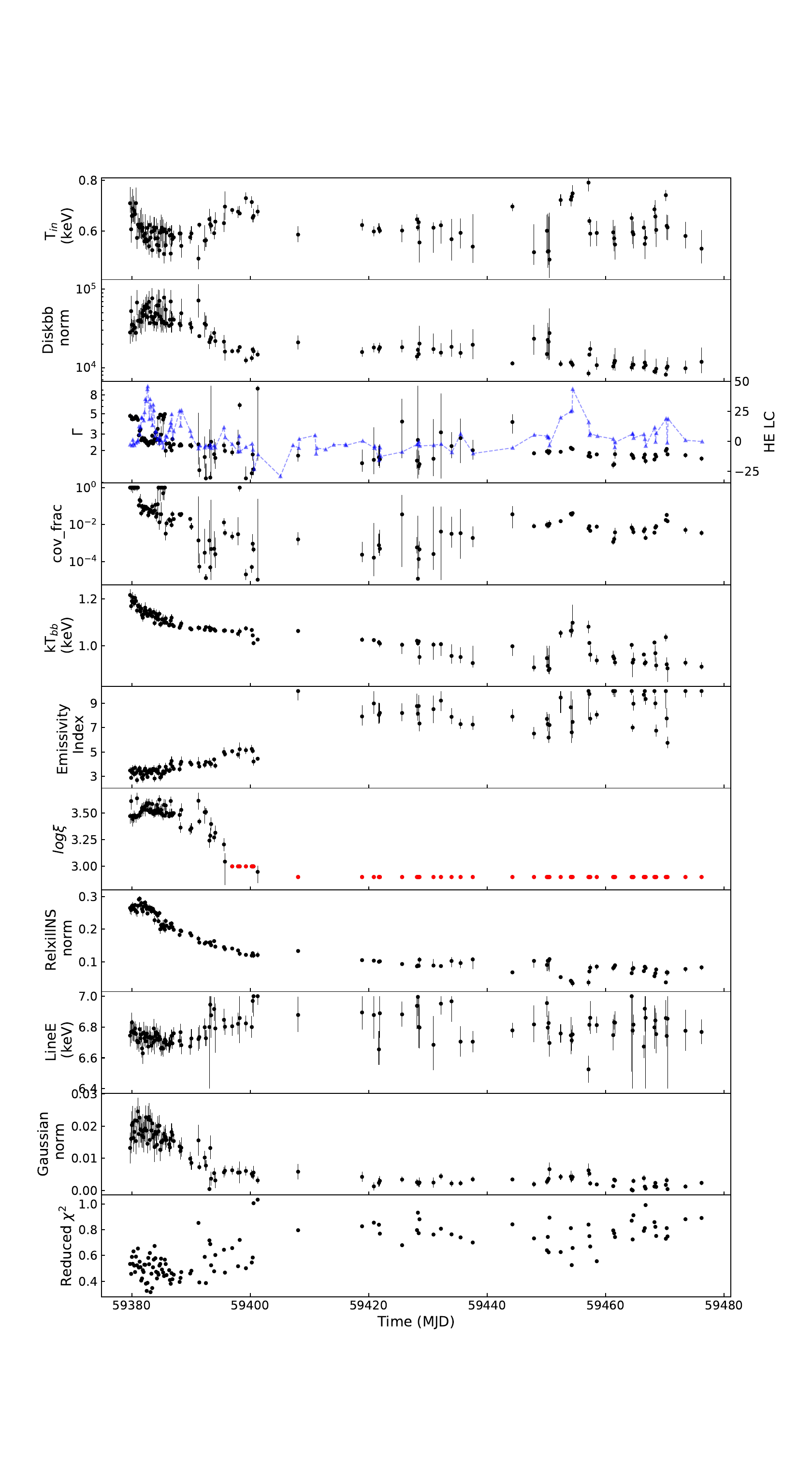}
	\vspace*{-30mm}
    \caption{Evolution of the spectral parameters in model \textrm{I} of 4U~1543$-$47 observed with \textit{Insight}-HXMT. From top to bottom, the panels show $T_{\rm in}$ and the normalization of the \textsc{diskbb}, $\Gamma$ and the covering fraction of \textsc{thcomp}, $kT_{\rm bb}$, the emissivity index, ionization of the accretion disk, log$\xi$, and the normalization of \textsc{relxillns}, the line energy and the normalization of the \textsc{gaussian} and the $\chi^2/{\rm d.o.f.}$. The red points represent parameters fixed.  The blue triangles in the third panel show HE photon count rate of 28.0--60.0 keV.}
    \label{fig:fig5_relxillns}
\end{figure*}

During our fits, the hydrogen column density, $N_{\rm H}$, has been fixed to $0.4\times 10^{22}$ cm$^{-2}$~\citep{2004ApJ...610..378P}.
In the \textsc{relxillns} component, we set the emissivity indices as q2$=$q1, so the intensity radial profile is described by a powerlaw. 
During the outburst, we assume the black-hole spin parameter, $a$, does not change and fix it to 0.67 which was obtained by~\citet{2020MNRAS.493.4409D}.
The inner radius of the accretion disk, $R_{\rm in}$, is fixed at the ISCO, because it always pegs to the lower limit in our initial attempt.
We fix the iron abundance, $A_{\rm Fe}$, to 1.0 times the solar value.
In the initial attempt, the density of the disk always pegs to the upper limit, $10^{19}$~cm$^{-3}$, at high luminosity, and it is not constrained at low luminosity. 
Recently~\citet{2021ApJ...909..146C} studied the X-ray spectra of BHB 4U 1630$-$47 by fitting a reflection model with variable disk density, and they found that the disk density is above $10^{20}$~cm$^{-3}$ across all spectral states. They concluded that the disk density in BHBs should be well above the standard value assumed in traditional reflection models (i.e., $\sim 10^{15}$~cm$^{-3}$).
We then fix the density to $10^{19}$~cm$^{-3}$, the upper limit in \textsc{relxillns}.
The inclination angle varied between 25$^{\circ}$ and 45$^{\circ}$ during the initial fits. Assuming a constant inclination angle, we then fix it to 35$^{\circ}$ during the outburst. 
We fix the reflection fraction to -1 so that this component only returns the reflected part of the spectrum and add a blackbody component outside the \textsc{relxillns} to account for the direct emission of the incident source, but the normalization of the blackbody is too small to contribute to the initial fits.
It is hard for \textit{Insight}-HXMT data to constrain the \textsc{gaussian} component, we therefore let the center line energy and normalization free and fix the line width to 0.2 in the fitting procedure.
We show the evolution of the parameters in Figure~\ref{fig:fig5_relxillns} and record them in Table~\ref{tab:the spectral parameters}.

In the first and second panels of Figure~\ref{fig:fig5_relxillns} we show the temperature at the inner disk radius, $T_{\rm in}$, and the normalization (defined as $(R_{in}/D_{10})^2\cos{i}$ where $R_{in}$ is an “apparent” inner disk radius in km, $D_{10}$ is the distance to the source in units of 10 kpc and $i$ is the inclination of the disk to the line of sight.) of the multicolor disk blackbody component. 
During the outburst decay, the temperature $T_{\rm in}$ varies between 0.5--0.7 keV and the normalization of \textsc{diskbb} decreases from $\sim5\times 10^4$ to $\sim1\times 10^4$. 
As shown in the fifth panel of Figure~\ref{fig:fig5_relxillns} The temperature of the black-body incident spectrum in \textsc{relxillns}, $kT_{\rm bb}$ decreases from $\sim$1.2 keV to $\sim0.9$ keV.

Since the X-ray spectra are dominated by the thermal component in all observations, we fix the corona temperature, $kT_{\rm e}$, of \textsc{thcomp} to 20 keV.
In the early decay phase (before MJD$\sim59400$), the power-law photon index, $\Gamma$, shows a large value of around 4.5 when the source is in the time intervals between the hard flares shown in the HE and ME light curves. This spectral state with high intensity and large $\Gamma$ is similar to the ``hypersoft'' state that appeared in the BHB GRO~J1655$-$40~\citep{2015MNRAS.451..475U}. Those observations in our fitting attempt with $kT_{\rm e}$ free show a very low electron temperature, $kT_{\rm e} \sim 2.0$ keV.
During the middle phase of the outburst between MJD$\sim$59400 and 59445, the X-ray spectra are dominated by the background emission above 15 keV, and $\Gamma$ can't be well constrained.
After 59445 (MJD) the photon index $\Gamma$ decreases to $\sim 2.0$ and remains constant until the end of the observations.

The emissivity index of \textsc{relxillns} shows significant differences before and after MJD$\sim$59400. Before that, the emissivity index increases very slowly from 3 to 5. After MJD$\sim$59400, the emissivity index remains constant at around 8.0.
The ionization parameter of the accretion disk, $\log\xi$, also shows a clear evolution. At the start of the outburst $\log\xi$ is around 3.5 and then starts to decrease from $\sim$59390 to $\sim$59405 (MJD). The ionization parameter can not be constrained well in the late decay of the outburst where $\log\xi \sim$ 2.9 with large error bars and has apparent effects on other parameters, like normalization of \textsc{relxillns}. Therefore, during our spectral analysis, we fix $\log\xi$ = 2.9 after MJD 59405. We also fix $\log\xi$ = 3.0 between 59396 and 59405 (MJD), a value between 2.9 and 3.1, the last value of $\log\xi$ before MJD 59396.

\subsubsection{The evolution of parameters with model \textrm{II}}

\begin{figure*}
    \vspace{-40mm}
	\includegraphics[width=0.9\textwidth]{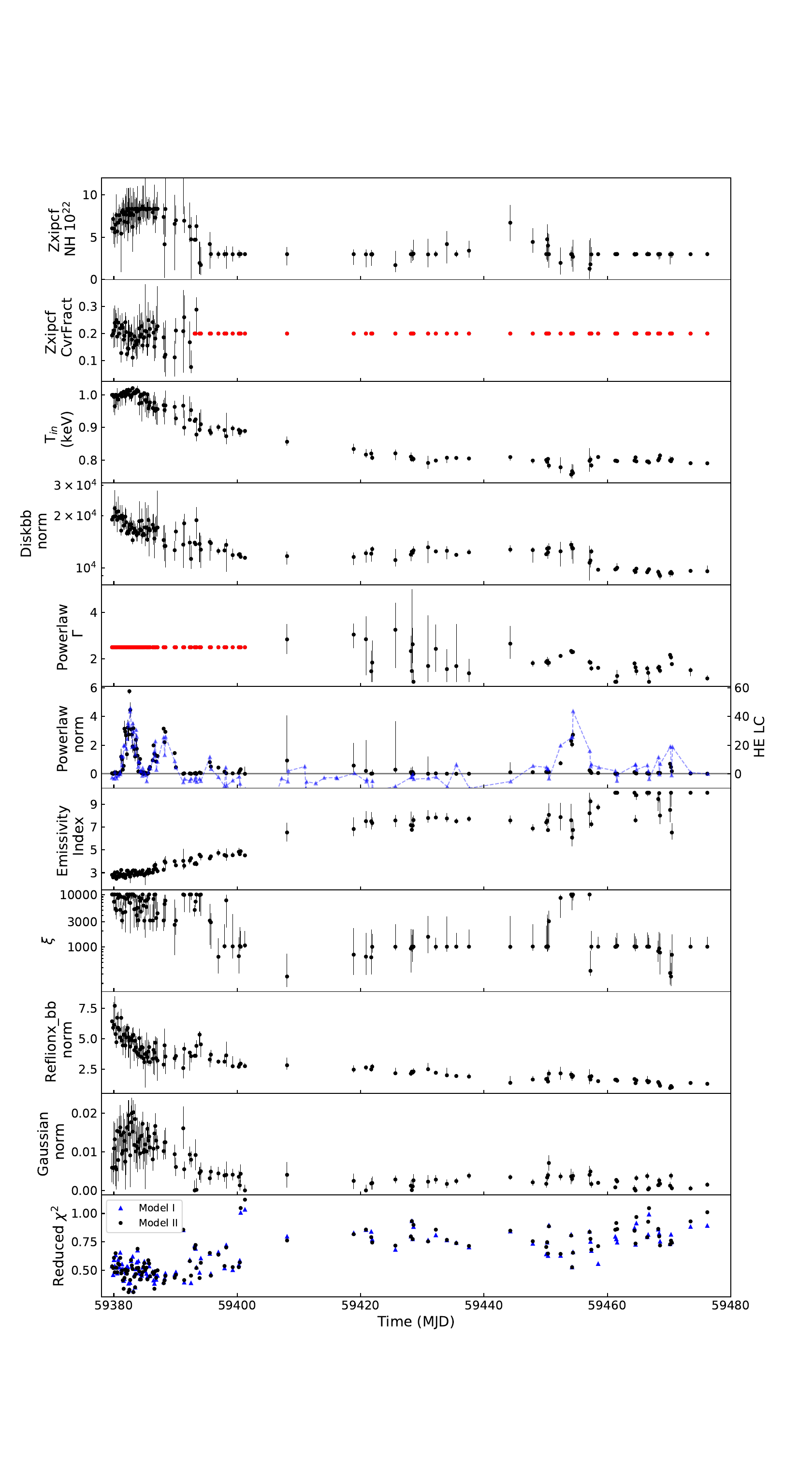}
	\vspace*{-30mm}
    \caption{ Evolution of the spectral parameters in model \textrm{II} of 4U~1543$-$47 observed with \textit{Insight}-HXMT. From top to bottom, the panels show nH and the covering fraction of \textsc{zxipcf}, $T_{\rm in}$ and the normalization of \textsc{diskbb}, photon index, $\Gamma$ and the normalization of \textsc{powerlaw}, the emissivity index of \textsc{relconv}, the ionization of the accretion disk, $\xi$, and the normalization of \textsc{reflionx\_bb}, the normalization of \textsc{gaussian} and the $\chi^2/{\rm d.o.f.}$. The red points represent parameters fixed. The blue triangles in the sixth panel show HE photon count rate of 28.0--60.0 keV. The blue triangles in the last panel show the $\chi^2/{\rm d.o.f.}$ of model \textrm{II}.}
    \label{fig:fig5_reflionx}
\end{figure*}

During our initial fitting, the ionization parameter in \textsc{zxipcf} is around 1.0 with large error bar.
As a result, the absorber describes the partial covering absorption by the low ionised medium.
The existence of the absorber during the decay of the outburst is also suggested by the near-infrared spectroscopy~\citep{2023A&A...673A.104S}.
We keep the ionization parameter in \textsc{zxipcf} fixed at 1.0 to reduce the degree of freedom.
The parameters in the \textsc{relconv} have been set the same as in \textsc{relxillns}, except for the inclination angle of the accretion disk that is fixed to 30$^{\circ}$ based on initial fitting. In the \textsc{reflionx\_bb} we link $kT_{\rm bb}$ to $T_{\rm in}$ of \textsc{diskbb}. Similar to the ionization parameter in \textsc{zxipcf}, the iron abundance, $A_{\rm Fe}$, is around 3.0 with larger error bar during our initial fitting.
So, we fix the iron abundance, $A_{\rm Fe}$, at 3.0. 
The density of the accretion disk is also fixed to the maximum value, $10^{20}$~cm$^{-3}$.
In the \textsc{gaussian} component, the center line energy and line width are fixed to 6.7 keV and 0.2 keV, respectively.
We show the evolution of the parameters in Figure~\ref{fig:fig5_reflionx} and record them in Table~\ref{tab:the spectral parameters_reflionx}.

The column density and covering fraction of \textsc{zxipcf} are shown in the first and second panel of Figure~\ref{fig:fig5_reflionx}, respectively.
The column density is relatively stable, ranging between 5 and 10 in the early decay phase and then decreases to 3.
The covering fraction firstly stays around 0.2 but is difficult to constrain during the later decay of the outburst. As a result, we chose to fix the covering fraction at 0.2 in the later stages of the decay.
The temperature of \textsc{diskbb} decreases from approximately 1.0 keV to around 0.8 keV, while the normalization decreases from roughly 20,000 to about 10,000.

\begin{figure}
    \vspace{-5mm}
	\includegraphics[width=\columnwidth]{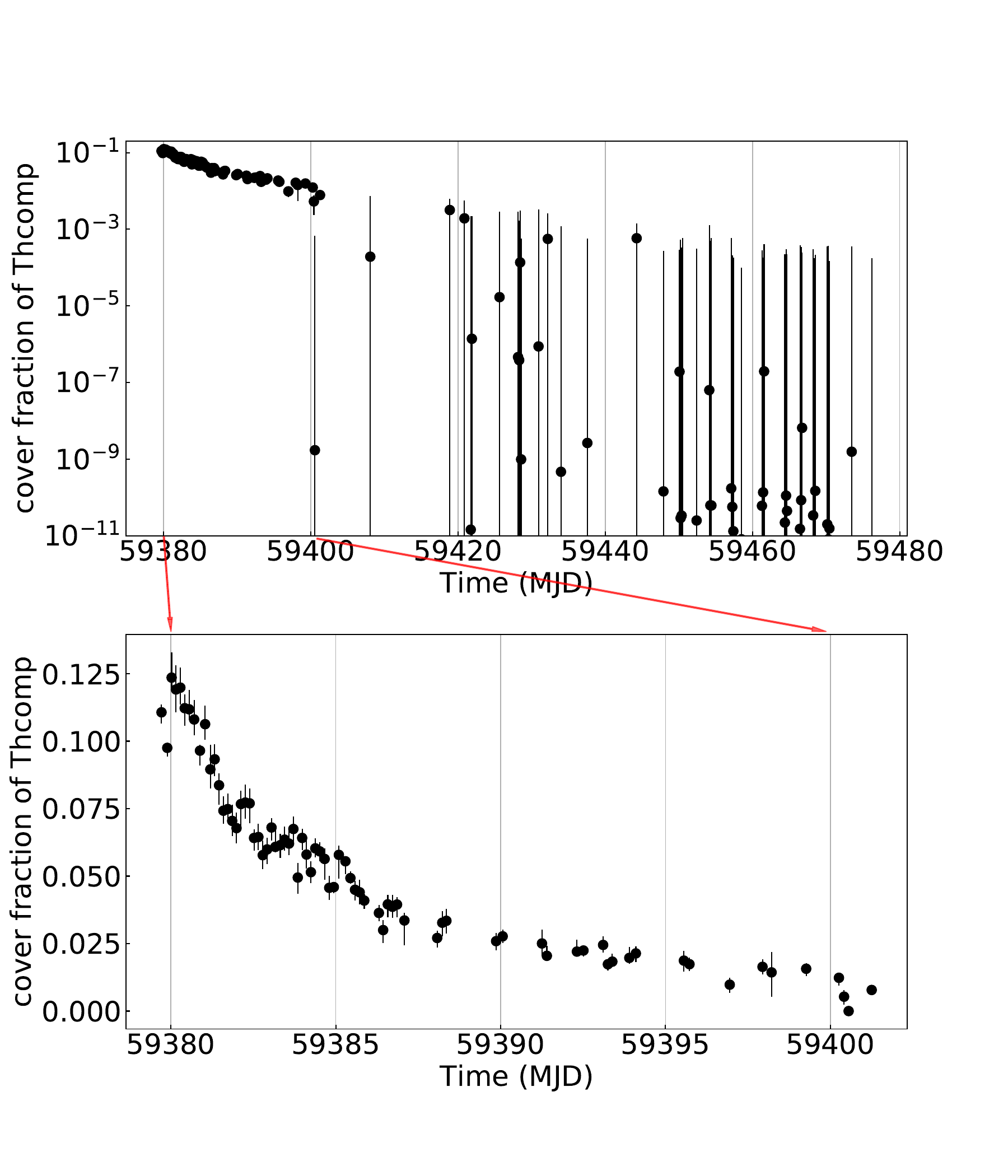}
	\vspace{-5mm}
    \caption{
    Top panel: the covering fraction of \textsc{thcomp}. Based on this figure, \textsc{thcomp} is removed after MJD 59405. 
    Bottom panel: the covering fraction of \textsc{thcomp} before MJD$\sim$59400.}
    \label{fig:thcomp_frac}
\end{figure}

We fix the electron temperature of \textsc{thcomp} at 2.0 keV based on the initial fitting.
Consequently, the warm Comptonized component seems to be similar to the soft excess presented in ULXs~\citep{2009MNRAS.397.1836G, 2021AstBu..76....6F} that usually are observed above the Eddington limit.
We also set the photon indices of both \textsc{thcomp} and \textsc{powerlaw} to 2.5 based on the initial fitting.
The top panel of Figure~\ref{fig:thcomp_frac} displays the covering fraction of \textsc{thcomp}, which drops below 0.0001 with a significant error after MJD$\sim$59400, indicating that \textsc{thcomp} is no longer necessary for observations in the later stages of the decay.
Consequently, we remove \textsc{thcomp} from MJD 59405.
To provide a closer look at the data before MJD$\sim$59400, we present a zoomed-in version in the bottom panel of the Figure. The covering fraction of \textsc{thcomp} decreases from 0.125 to 0.005 before MJD$\sim$59400.
The photon index and normalization of \textsc{powerlaw} are displayed in the fifth and sixth panel of Figure~\ref{fig:fig5_reflionx}, respectively.
Notably, the normalization of \textsc{powerlaw} traces the HE 28.0--60.0 keV light curve, as displayed in the sixth panel with blue points. 
Although the \textsc{powerlaw} component is added to all observations, it is not really needed for the observations in which the source is not detected in the HE spectrum. For example, as mentioned at the penultimate paragraph of the section~\ref{sec:Model selection}, for observations outside of the hard flares before MJD 59400, \textsc{thcomp} with kT$_{\rm e} \sim$2.0 keV is enough.

In model \textrm{I} we only add a Comptonized component \textsc{thcomp}. Notably, before MJD 59400, the photon index ($\Gamma$ of \textsc{thcomp} in model \textrm{I}) evolves discontinuously. Outside of the hard flares, a \textsc{thcomp} component with a large photon index describes the soft excess. During the hard flares, a \textsc{thcomp} component with a low photon index tends to describe the hard photons produced by the hard flares, instead of the soft excess.
As a result, the evolution of the parameters of \textsc{thcomp} in model \textrm{I} are discontinuous.

The emissivity index of \textsc{relconv} also shows significant differences before and after MJD$\sim$59400, same as in \textsc{relxillns}. Before MID 59400 the emissivity index increases very slowly from 3 to 5. However, after MJD$\sim$59400 the emissivity index remains at around 7--9.
The ionization parameter of the accretion disk, log$\xi$, shows a decrease $\sim$4 to $\sim$3.

In the bottom panel of Figure~\ref{fig:fig5_reflionx}, we show the $\chi^2/{\rm d.o.f.}$ with black points. As a comparison, the $\chi^2/{\rm d.o.f.}$ with \textsc{relxillns} are also shown with blue points. It is evident that the $\chi^2/{\rm d.o.f.}$ values from both reflection models exhibit a similar trend.

\begin{figure}
    \vspace{-5mm}
	\includegraphics[width=\columnwidth]{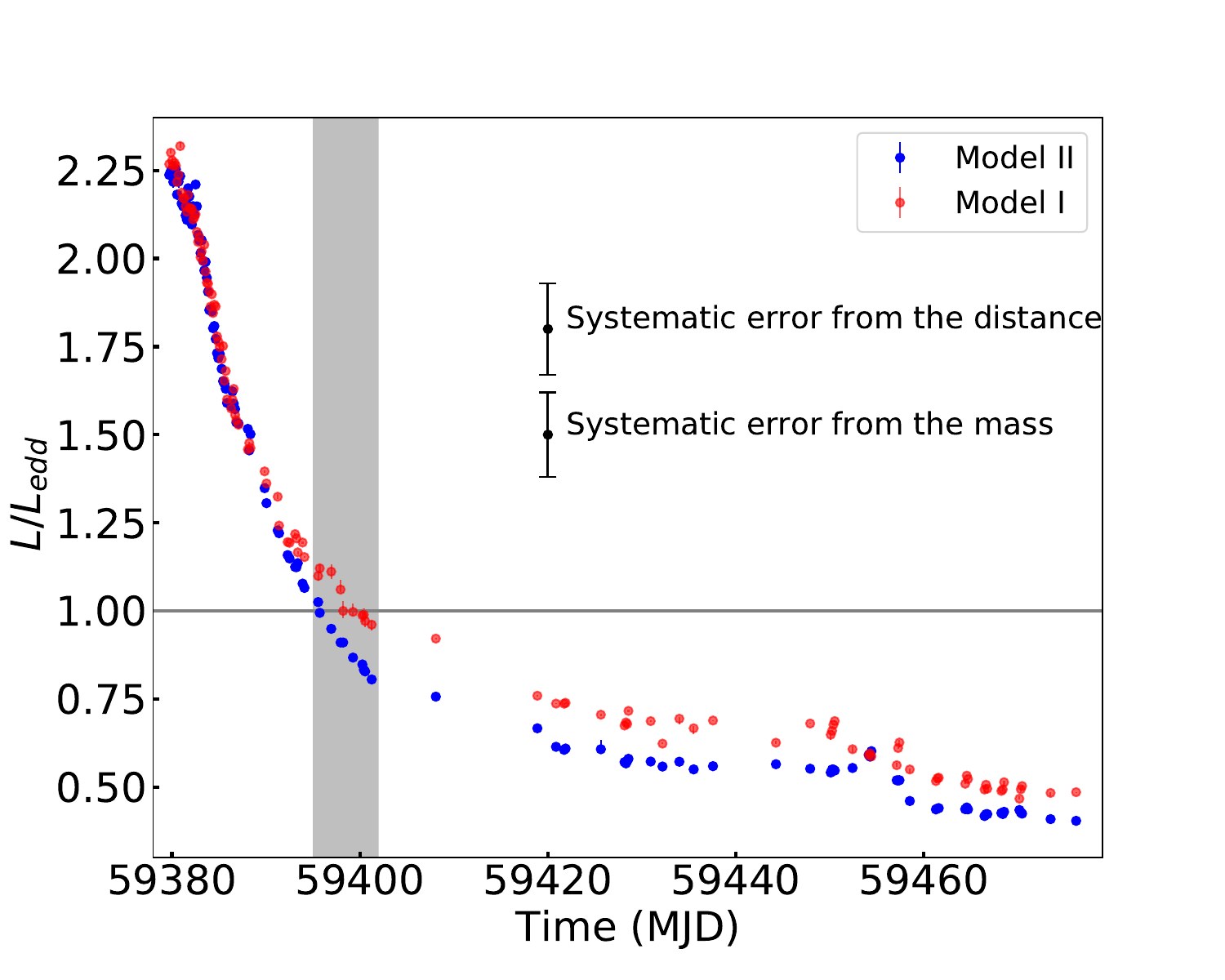}\vspace{-8mm}
        \includegraphics[width=\columnwidth]{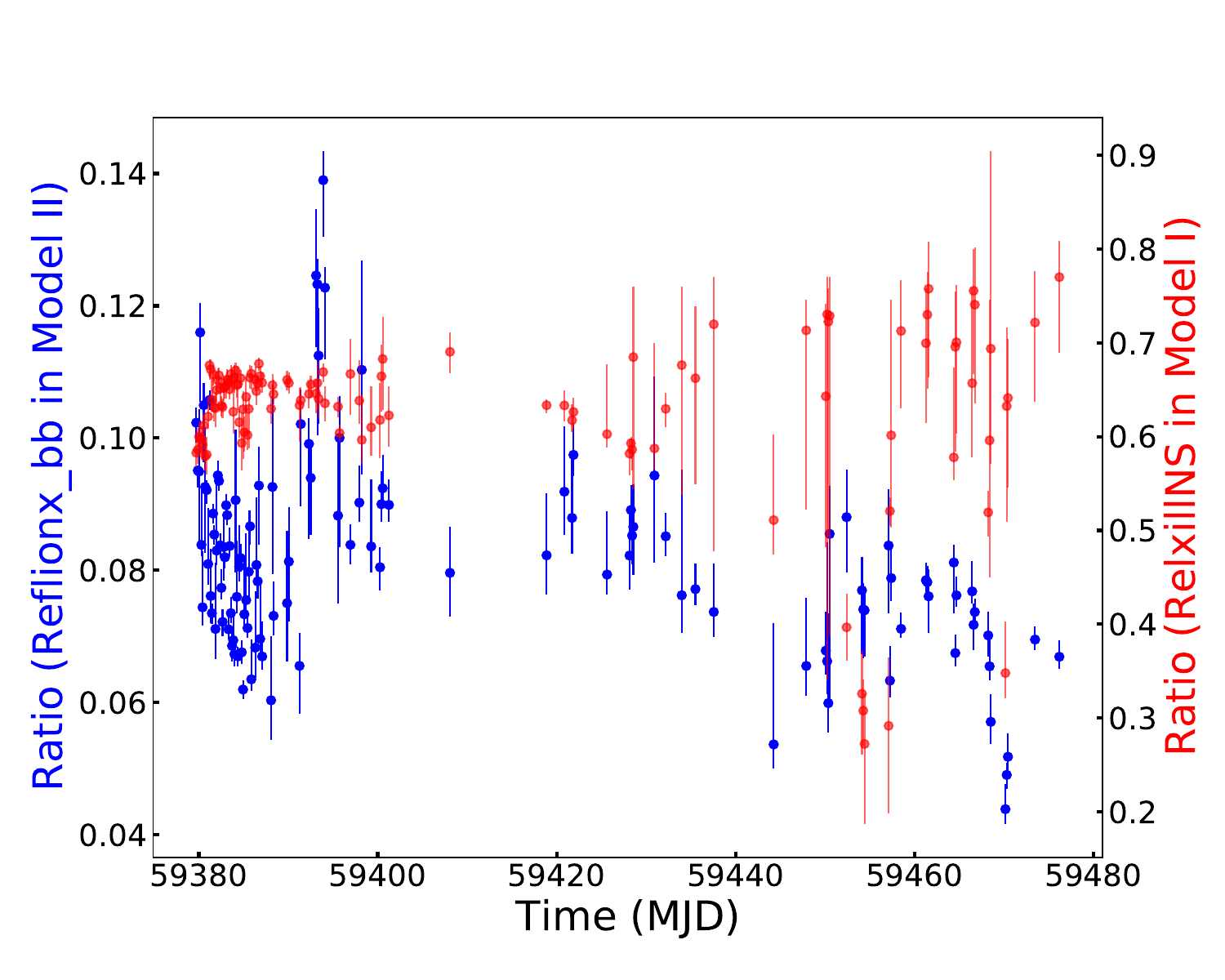}
	\vspace{-8mm}
    \caption{Top panel: The evolution of the luminosity obtained by the both of reflection models, \textsc{relxillns} and \textsc{reflionx\_bb}, of 4U 1543$-$47 in Eddington units. The points in the gray region correspond to the observations in the gray region of Figure~\ref{fig:fig_hid_fit}. 
    We also show the systematic errors due to the mass, 9.4 $\pm$ 1.0 $M_\odot$~\citep{2003A&A...404..301R}, and the distance, 7.5 $\pm$ 0.5 kpc~\citep{2002AAS...201.1511O, 2004MNRAS.354..355J}, respectively.
    Bottom panel: The ratio of the flux of the reflection components to the total flux.}
    \label{fig:flux}
\end{figure}

\subsubsection{Flux obtained from Models \textrm{I} and \textrm{II}}

We calculate the total unabsorbed flux in the 0.1--60 keV energy band and show the light curve in units of the Eddington luminosity in the top panel of Figure~\ref{fig:flux}. In calculating flux in Eddington units, we use a black hole mass of 9.4 $M_\odot$~\citep{2003A&A...404..301R} and a distance of 7.5 kpc~\citep{2002AAS...201.1511O, 2004MNRAS.354..355J}.
Compared with the MAXI light curve, the first \textit{Insight}-HXMT observation takes place at the peak of the outburst. 
The peak luminosity of 4U~1543$-$47 during the 2021 outburst is $\sim 2.3 L_{\rm Edd}$, regardless of which reflection model is used, much brighter than the outburst peak luminosity in other BHBs.
After MJD$\sim$59400, the luminosity obtained using \textsc{relxillns} (model \textrm{I}) is slightly higher than that with \textsc{reflionx\_bb} (model \textrm{II}). This difference could be attributed to the \textsc{relxillns} model overestimating the flux below 2 keV, as this model exhibits a flat spectrum at low energies (For more details, refer to section~\ref{sec:The super-Eddington state}).

In the bottom panel of Figure~\ref{fig:flux} we present the ratio between the reflection component and the total flux during the outburst for \textsc{relxillns} (in red) and \textsc{reflionx\_bb} (in blue). 
In model \textrm{I} we obtain that more than 50\% of the flux comes from the reflection component for almost all observations. This value remains constant during the whole outburst and does not show a clear evolution. 
The ratio of the reflection component in model \textrm{I} is significantly higher than expected in the study of returning disk radiation~\citep{2005ApJS..157..335L, 2022MNRAS.514.3965D}.
On the contrary, in model \textrm{II} we obtain that less than 10\% of the flux is due to reflection,
the contribution of \textsc{reflionx\_bb} is consistent with the predictions of returning disk radiation~\citep{2005ApJS..157..335L, 2022MNRAS.514.3965D}.

\subsection{The State Transition near the Eddington Luminosity}
\label{sec:The State Transition near One Eddington Luminosity}

\begin{figure}
    \vspace{-10mm}
	\includegraphics[width=\columnwidth]{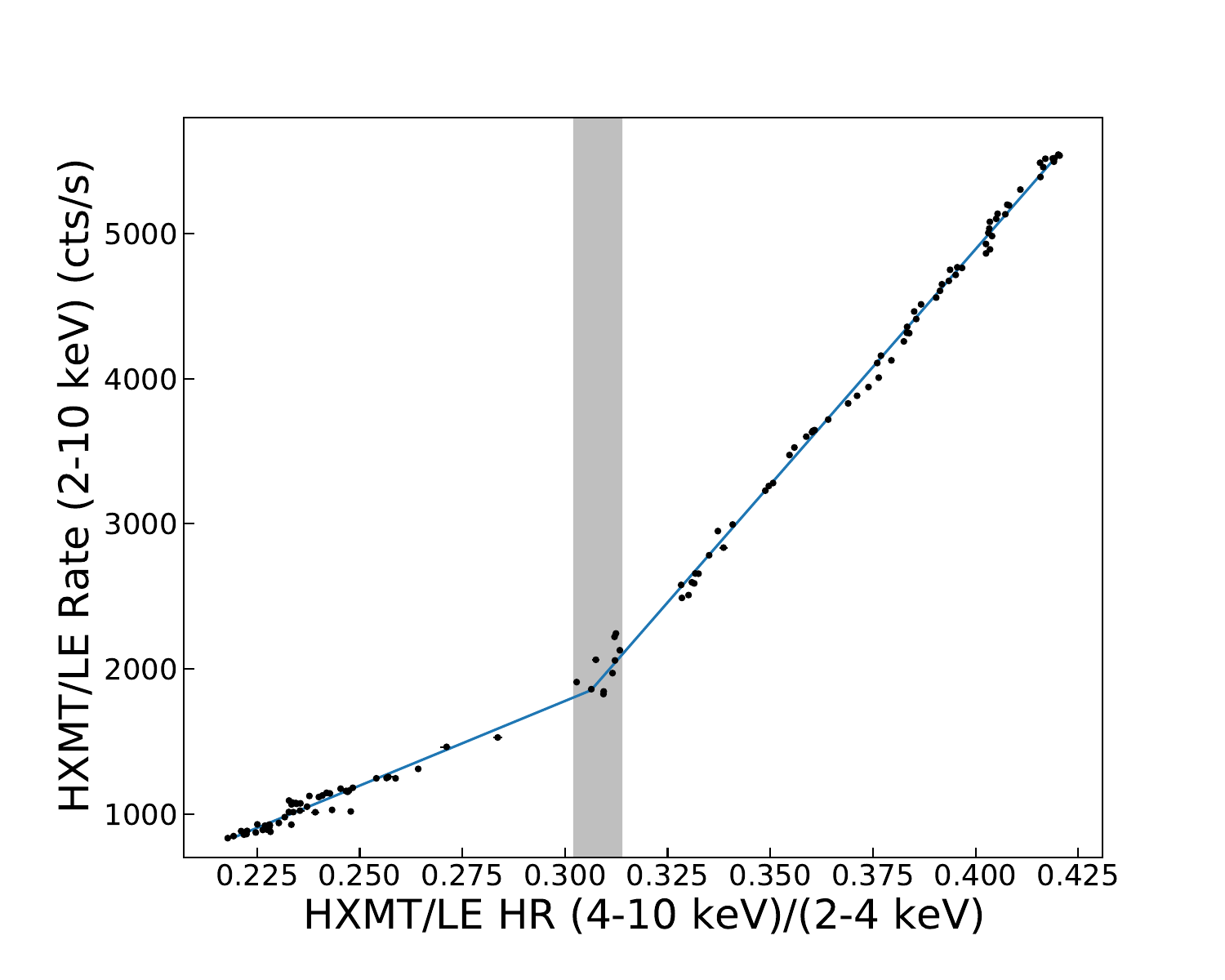}\vspace{-8mm}
	\includegraphics[width=\columnwidth]{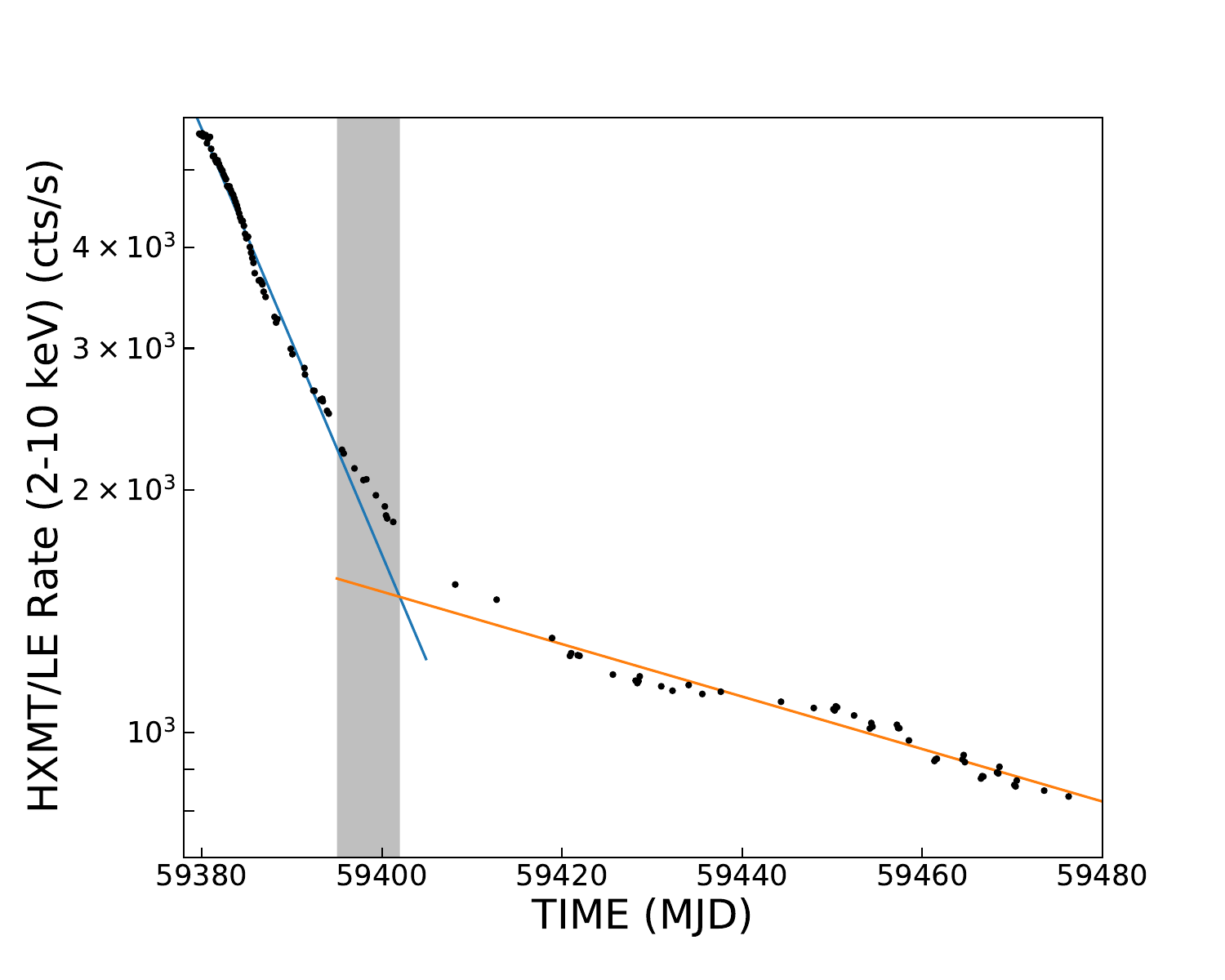}
	\vspace{-8mm}
    \caption{Top panel: HID of 4U 1543$-$47 from \textit{Insight}-HXMT/LE (4--10 keV)/(2--4 keV), fitted with a broken line. Bottom panel: The 2--10 keV light curve of 4U 1543$-$47, fitted with two exponential functions for the periods MJD $<$ 59395 and $>$ 59405.}
    \label{fig:fig_hid_fit}
\end{figure}

As it is apparent in the \textit{Insight}-HXMT HID (Figure~\ref{fig:HID}), there are two branches in the diagram, indicating a state transition. In order to investigate the location of the spectral transition, in the top panel of Figure~\ref{fig:fig_hid_fit} we fit the data in the \textit{Insight}-HXMT HID with a broken-line function. 
The parameters are the hardness ratio, HR$_{\rm bk}$, and intensity, $I_{\rm bk}$, at the break, and $a_1$ and $a_2$, the slopes at the high and low-intensity intervals, respectively. Because the scatter of points is far greater than the statistical errors, we use the least squares method to estimate the parameter errors. The goodness of the fit, $\rm 1-(\Sigma (y-\hat{y})^2)/(\Sigma (y-\bar{y})^2)$, is 0.999.
We get HR$_{\rm bk} = 0.306 \pm 0.001$ and $I_{\rm bk} = 1850 \pm 60$ cts/s. 
The data around the break are marked by the gray region.
The slopes $a_1$ and $a_2$ are 32500 $\pm 200$ counts/s per unit hardness and 11700 $\pm 500$ counts/s per unit hardness, respectively. The slope $a_1$ during the high-intensity interval is $\sim$ 2.8 times $a_2$, indicating that the source spectra in the early decay phase evolve faster than that in the late decay.

In most of the outburst light curves in BHBs there are several periods of exponential decay as the source evolves in different spectral states~\citep[e.g., ][]{2015MNRAS.450.3410E} or accretion regimes~\citep{1998MNRAS.301..382S}. To study the decay time scale in different spectral states, we divide the LE 2--10 keV light curve into two periods based on the HID fits above, before the state transition marked by the gray region (top panel of Figure~\ref{fig:fig_hid_fit}, MJD $<$ 59395), and after the transition. We then fit the light curves of the two periods with an exponential function, $A\exp(-t/\tau)$. We show the light curve fits in the bottom panel of Figure~\ref{fig:fig_hid_fit}. The points in the gray region are the same observations marked in the gray region in the HID. The two best-fitting exponential decay curves meet at MJD $\sim 59402$, just at the edge of the gray region.
We get a decay time-scale $\tau_{1} = 16.4 \pm 0.2$ days in the first period and $\tau_{2} = 133 \pm 5$ days in the second period. The decay time-scale in the second period is $\sim 8$ times that in the first period. The time scale in the second period implies that the accretion rate remains at a high level for a long time during the outburst.

The above analysis of the HID indicates that there is a clear state transition during the outburst decay.
We also mark the data during the state transition with the gray region in the light curve in Eddington units (top panel of Figure~\ref{fig:flux}). Interestingly, the source luminosity during this transition is close to $1 L_{\rm Edd}$.
The light curve can be divided into two periods: super-Eddington and sub-Eddington, with the change from one to the other happening at the time around the state transition.

\begin{figure}
\vspace{-8mm}
	\includegraphics[width=\columnwidth]{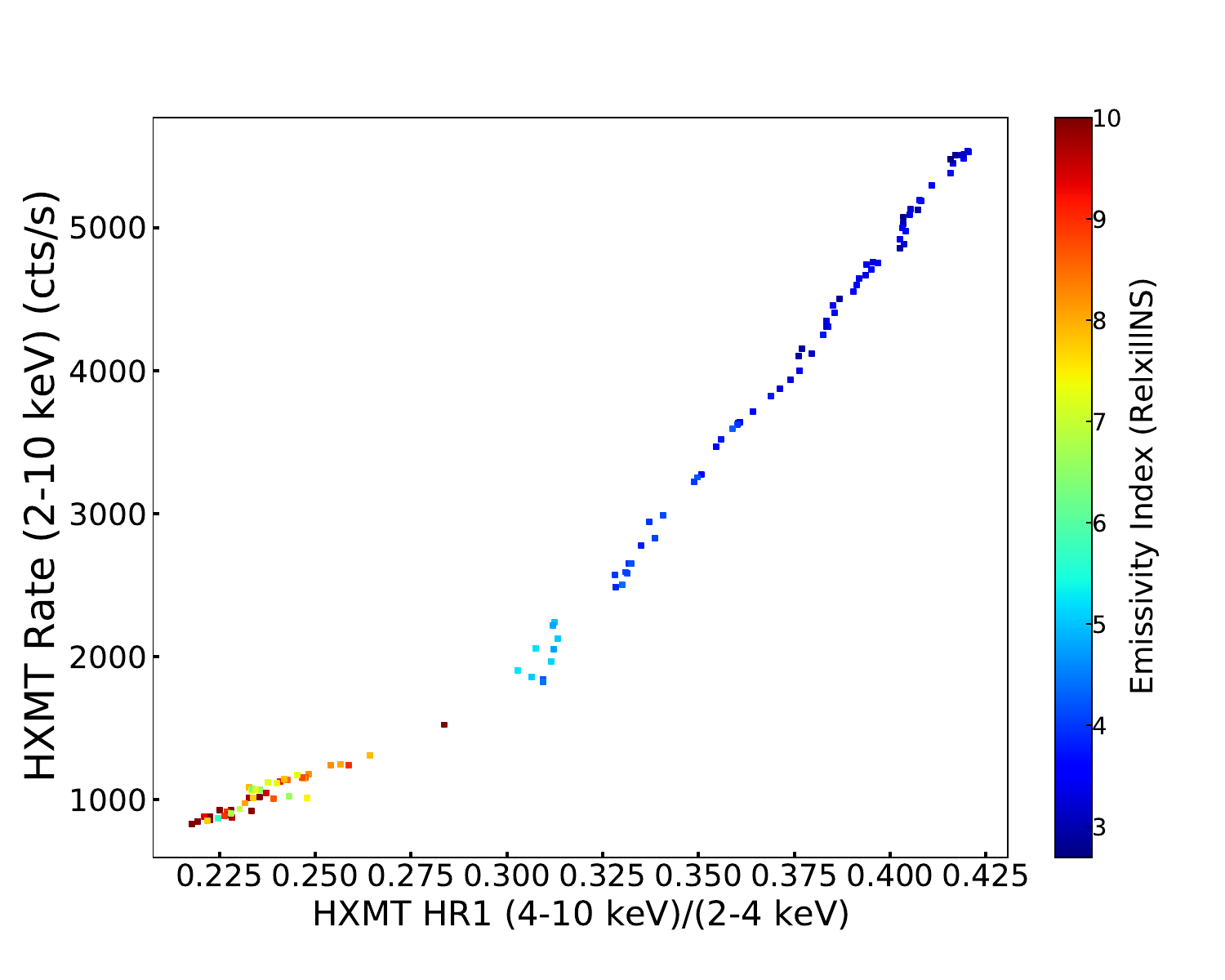}\vspace{-8mm}
        \includegraphics[width=\columnwidth]{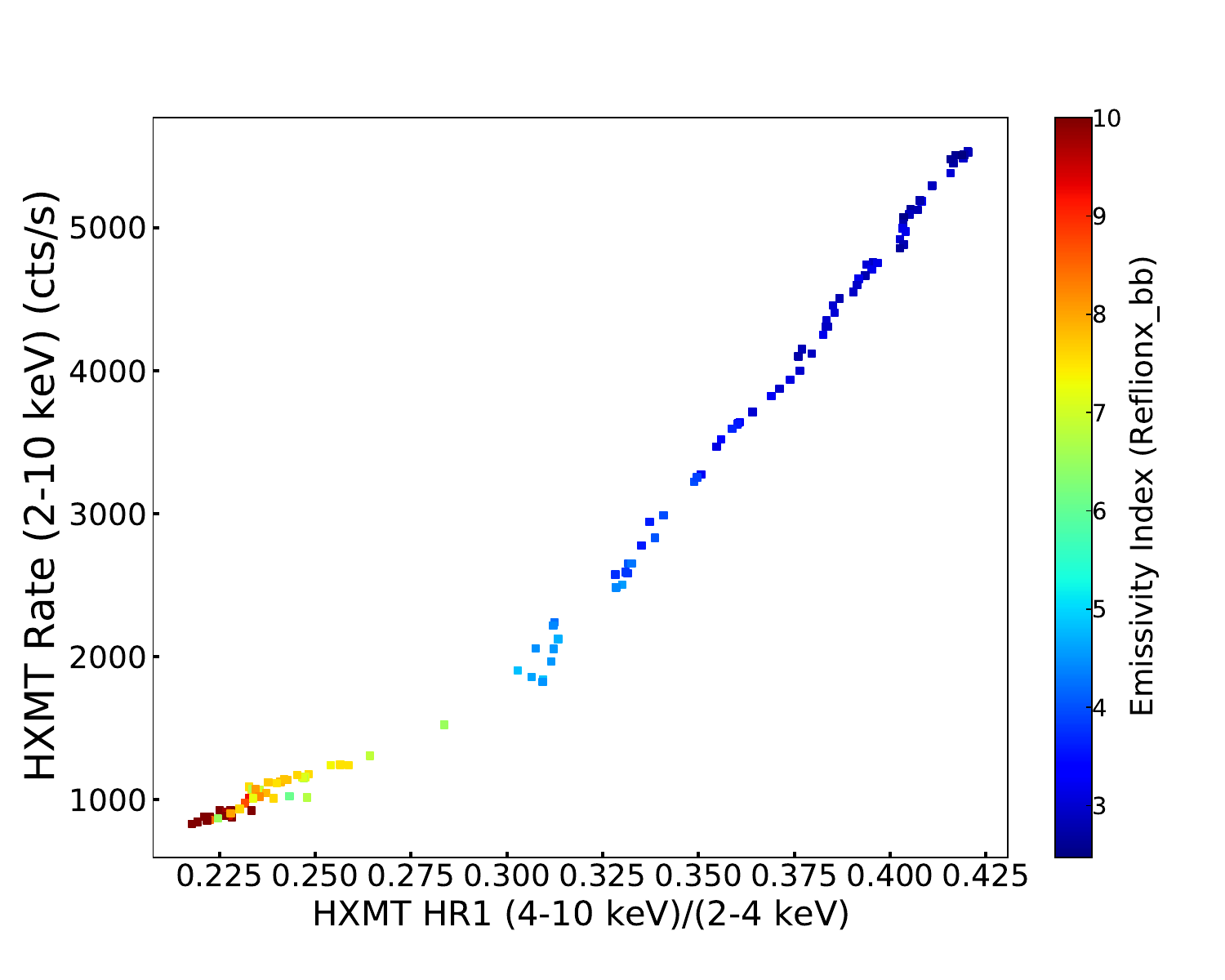}
 \vspace{-8mm}
    \caption{The HID of 4U~1543$-$47 observed by \textit{Insight}-HXMT/LE. The colors of the points represent the emissivity index, top panel for \textsc{relxillns} and bottom panel for \textsc{reflionx\_bb}, respectively.}
    \label{fig:fig_index_incl}
\end{figure}

We have already mentioned that the emissivity index shows significant differences before and after MJD$\sim$59400, close to the transition, regardless of which reflection model we use. In Figure~\ref{fig:fig_index_incl} it is apparent that the emissivity index shows a clear difference between the super- and sub-Eddington branches in the HID.
In addition, a Comptonized component with low electron temperature $\sim$2.0 keV appears before MJD$\sim$50400 or the state transition and disappears after the state transition(see Figure~\ref{fig:thcomp_frac}).
The luminosity-temperature (L-T) relation of the outburst had been studied by~\citet{2023A&A...677A..79Y}.
They found that the L-T relation also changes clearly during the state transition.

\section{DISCUSSION}
\label{sec:discussion}

We have analyzed 121 time-averaged X-ray spectra of the BHB 4U~1543$-$47 observed with Insight-\textit{Insight}-HXMT during the 2021 outburst. The \textit{Insight}-HXMT spectra are dominated by the soft photons from the disk, with a strong broad iron emission line accompanied with a hump at 10--20 keV. 
The 2--60 keV broad-band X-ray spectra can be well fitted with reflection models with a black-body incident spectrum. 
The peak luminosity of the outburst is $\sim 2.8\times 10^{39}$~erg s$^{-1}$, or $\sim$2.3 $L_{\rm Edd}$, making it the most luminous outburst in recent years~\citep{2021ATel14708....1N}.
We find a clear state transition of this source from the HID analysis.
Interestingly, the source luminosity during this transition is close to $\sim 1 L_{\rm Edd}$.
The 2.0--10.0 keV LE light curve shows different decay time scales before and after this transition.
The emissivity index changes clearly around the transition. The Comptonized component with low electron temperature appears only before the state transition.

\subsection{Comparison with Earlier Outbursts of 4U~1543$-$47}
\label{sec:Comparison with Earlier Outbursts}

In Table ~\ref{tab:outburst} we show the years of each outburst, the time interval between outbursts and the peak intensity of the outbursts of 4U~1543$-$47 that have been reported.
Since 4U~1543$-$47 was first discovered in 1971~\citep{1972ApJ...174L..53M}, the source was in outburst in  1971, 1983, 1992, and 2002~\citep{2004ApJ...610..378P}. 
The time intervals of quiescent states between previous outbursts are $\sim 10$ years, but the recurrence time for the 2021 outburst is about 20 years, twice the usual interval.
Comparing the distribution of the outburst recurrence times for the Galactic transient BHBs population between 1996 and 2015 in Figure 10 of~\citet{2016ApJS..222...15T}, the $\sim 20$-year recurrence time in 4U~1543$-$47 is longer than the recurrence time in most of the outbursts detected in other BHCs. 
The very long recurrence time may be one of the reasons that the 2021 outburst is the brightest outburst of this source so far, about 10 Crab (\text{MAXI}/GRC 3.7--7.5 keV) at the maximum.
The peak intensities of earlier outbursts are 1.9 Crab (2--6 keV) in 1971, 4.0 Crab (3.7--7.5 keV) in 1983, and 3.3 Crab (3.7--7.5 keV) in 2002~\citep{2004ApJ...610..378P}.
The 2021 outburst shows a similar rise time as the 2002 outburst, $\sim 4$ days, but a much longer decay time than the 2002 outburst, which lasted only about 40 days~\citep[see also,][]{2004ApJ...615..880B, 2005ApJ...622..508K, 2020MNRAS.495..182R}. 
In addition, the long decay time-scale during the sub-Eddington period ($\sim 130$ days) also indicates that a large amount of accretion material was stored during the quiescent state for $\sim 20$ years.

\begin{table}
	\centering
	\caption{The year, the recurrence time and the peak intensity of the outbursts of the BHB 4U~1543$-$47 that have been reported. The peak intensities of earlier outbursts are from~\citep{2004ApJ...610..378P}.}
	\label{tab:outburst}
	\begin{tabular}{lccr}
		\hline
		Year & Recurrence Time (years) & Peak Intensity \\
		\hline
		1971 &  & 1.9 Crab (2--6 keV)\\
		1983 & 12 & 4.0 Crab (3.7--7.5 keV)\\
		1992 & 9 & -- \\
		2002 & 10 & 3.3 Crab (3.7--7.5 keV) \\
		2021 & 19 & $\sim$ 10 Crab (3.7--7.5 keV) \\
		\hline
	\end{tabular}
\end{table}

\subsection{Comparison with the 2005 Outburst of GRO~J1655$-$40}

The MAXI/GSC HID of the 2021 outburst of 4U~1543$-$47 does not show the common ``q''-shaped HID with the transition from LHS to HSS at maximum intensity. After the outburst starts, 4U~1543$-$47 first moves from the lower-right to the lower-left on the HID, implying that the source enters the soft state without a clear increase of the intensity. The HR then is positively correlated with intensity in both the rising and decay phases, and a ``diagonal bar'' shape appears in the HID at the peak.
As described in Section \ref{sec:introduction}, the branch at high intensity is similar to the reported states in the BHB GRO~J1655$-$40 when the outburst is very bright, where either the shape of the HID, the spectral or timing properties are clearly different from the HSS, such as the ``steep powerlaw'' state in the 1996 outburst~\citep{2006ARA&A..44...49R}, the ``ultra-luminous'' state~\citep{2012MNRAS.427..595M} and the ``hypersoft'' state~\citep{2015MNRAS.451..475U} in the 2005 outburst.

A similar ``diagonal bar'' shape at high intensity also appears in the HID of the 2005 outburst in BHB GRO~J1655$-$40, corresponding to the ``ultra-luminous'' state~\citep{2012MNRAS.427..595M}. However, the ``ultra-luminous'' state of GRO~J1655$-$40 shows higher hardness (the ratio of the 15--20 keV to the 3--5 keV flux) than its soft state, although the 15--20 keV photon index is the same as that of the soft state~\citep[$\sim$ 2.0--3.0;][]{2015MNRAS.451..475U}. This is different from 4U~1543$-$47 that showed a large photon index $\Gamma \sim 4.5$ during the super-Eddington period, which indicates that 4U~1543$-$47 is softer during the super-Eddington period than in the soft state (the sub-Eddington period). 
In addition, QPOs~\citep{2012MNRAS.427..595M} have been detected in the ``ultra-luminous'' state of GRO~J1655$-$40. However, we do not find any QPO signals between 0.01 Hz and 500 Hz in 4U~1543$-$47 (Jin et al. in preparation).
Therefore, we suggest that the super-Eddington branch in our analysis is different from the ``ultra-luminous'' state of GRO~J1655$-$40.

The large photon index during the super-Eddington period is more similar to the ``hypersoft'' state with photon index $> 5$ of GRO~J1655$-$40~\citep{2015MNRAS.451..475U}, softer than the soft state.
The large photon index may result from electrons with low temperature in the region of inverse-Compton scattering~\citep{1995ApJ...450..876T, 1980A&A....86..121S, 2016ApJ...822...20N}.
A comptonized component with low electron temperature (the soft excess, Figure~\ref{fig:thcomp_frac}) is also found during super-Eddington period of 4U~1543$-$47.
The Comptonized component vanishes as the source transitions to the sub-Eddington state.
In the ``hypersoft'' state, the black hole GRO~J1655$-$40 could be also at or above the Eddington limit~\citep{2015MNRAS.451..475U, 2016ApJ...822...20N}.

\subsection{The super-Eddington state}
\label{sec:The super-Eddington state}

Different models have been developed to describe accretion and radiative process at super- and sub-Eddington limits~\citep{1973A&A....24..337S, 2007MNRAS.377.1187P, 2017ARA&A..55..303K, 2021AstBu..76....6F}.
Due to the radiation pressure, the thickness of the accretion disk increases significantly near the Eddington limit~\citep[Figure 1 of][]{2016A&A...587A..13L}.
When the luminosity exceeds the Eddington limit, a supercritical accretion disk is expected~\citep{1973A&A....24..337S, 2007MNRAS.377.1187P, 2017ARA&A..55..303K, 2021AstBu..76....6F}.

In our analysis, we observe that the soft excess (Figure~\ref{fig:thcomp_frac}) appears in the super-Eddington state but disappears as the luminosity drops below the Eddington limit.
This soft excess is also consistent with that of the Comptonizing medium observed in ULXs~\citep{2009MNRAS.397.1836G, 2021AstBu..76....6F}.
Some effects, such as the MRI that drives large-scale turbulence~\citep{2004ApJ...601..405S}, could distort the spectrum of the accretion disk and give rise to the excess. This phenomenon is expected in the inner radiation-pressure dominated region of the accretion disk when the accretion rate reaches, or is above, the Eddington limit~\citep{2012MNRAS.420.1848D}. 
The decrease of the covering fraction of \textsc{thcomp} with decreasing luminosity (Figure~\ref{fig:thcomp_frac}) may imply that the radiation-pressure dominated region of the accretion disk shrinks and then disappears.

\begin{figure}
	\includegraphics[width=\columnwidth]{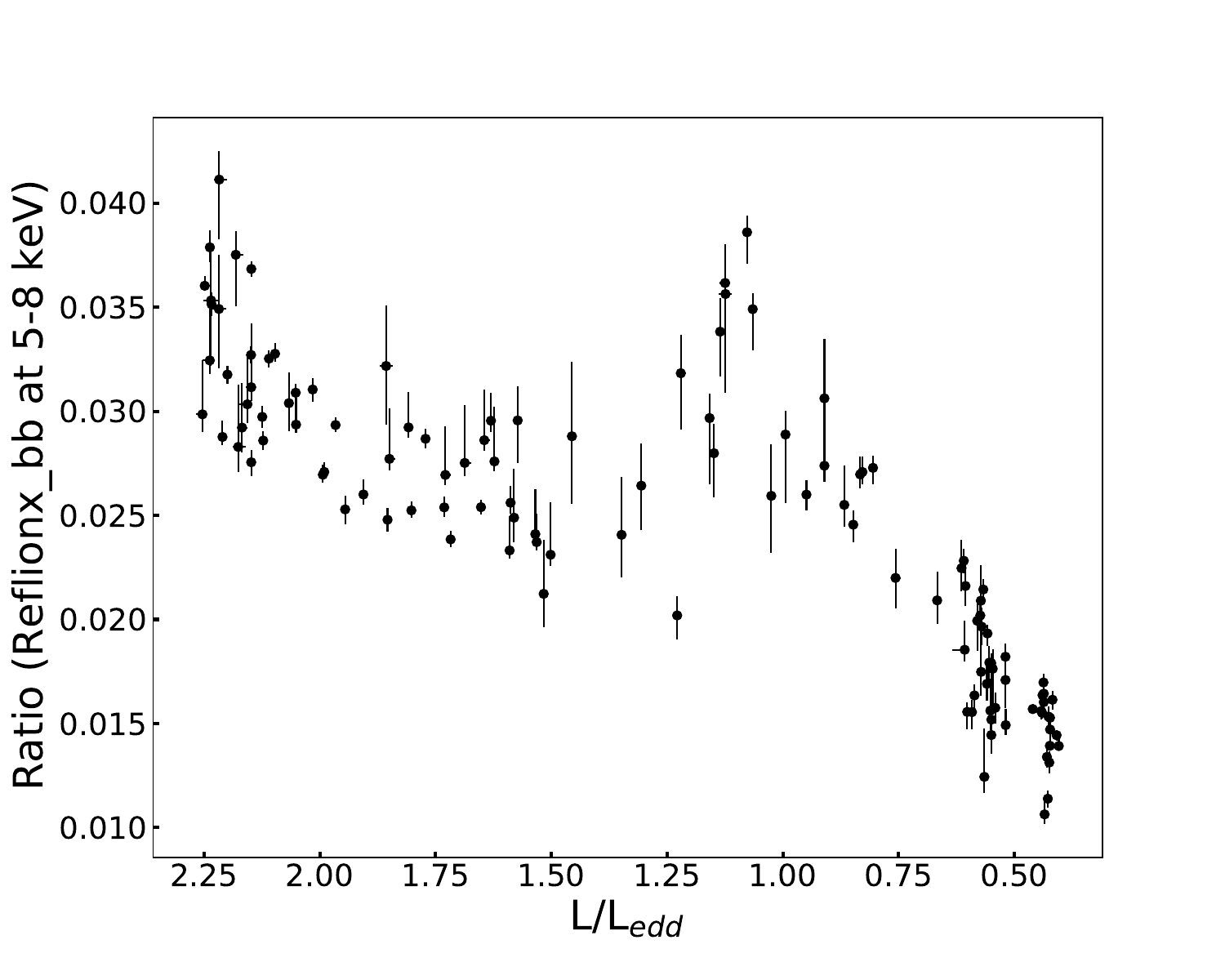}	
	\vspace{-5mm}
    \caption{The ratio of the reflection flux in the 5.0--8.0 keV band to the total flux in the 0.1--60.0 keV band. } 
    \label{fig:flux_ratio_5-8}
\end{figure}

At super-Eddington luminosities, a supercritical accretion disk funnel could make it easier for the accretion disk emission to illuminate the accretion disk itself.
Thus stronger reflection, especially the broad iron line at 5.0--8.0 keV, is expected at super-Eddington luminosities.
We plot the ratio of the reflection flux in the 5.0--8.0 keV band to the total flux in the 0.1--60.0 keV band in Figure~\ref{fig:flux_ratio_5-8}.
This ratio remains relatively stable, fluctuating within the range of 0.025 to 0.035 when the luminosity is above the Eddington, but it experiences a significant drop after the state transition.

\begin{figure}
	\includegraphics[width=\columnwidth]{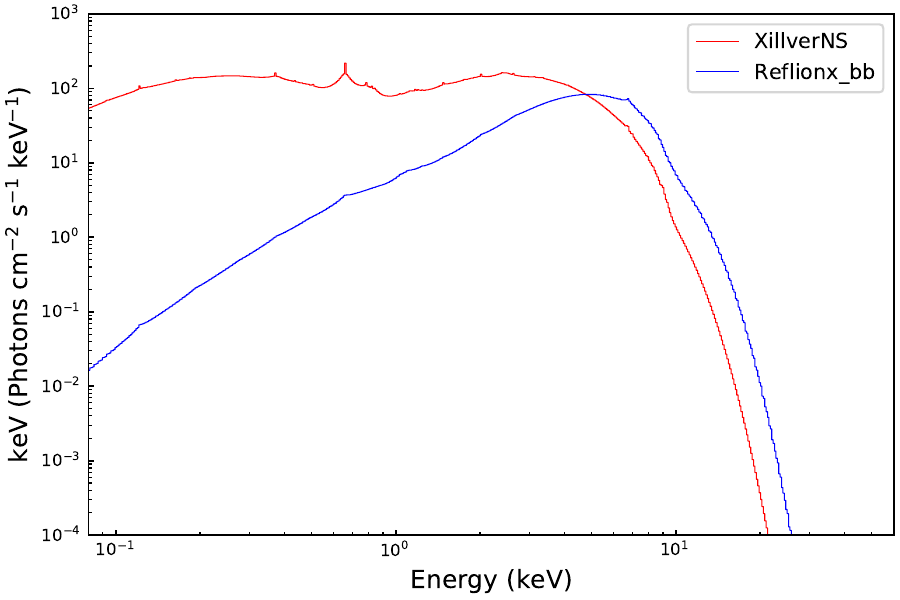}	
	\vspace{-5mm}
    \caption{The spectra of the reflection components \textsc{xillverns} and \textsc{reflionx\_bb} for an input black-body spectrum of 1.0 keV.
    The density of the accretion disk is $10^{19}$~cm$^{-3}$. The iron abundance is 1.0 and the ionization parameter is $\log\xi = 3.5$.
    The inclination of the disk in \textsc{xillverns} is 30$^{\circ}$.}
    \label{fig:flux_xillver_reflionx}
\end{figure}

In Figure~\ref{fig:flux_xillver_reflionx} we plot the reflected spectra obtained, respectively, by \textsc{xillverns} (non-relativistic version of \textsc{relxillns}) and \textsc{reflionx\_bb}, with an input black-body spectrum of 1.0 keV. The density of the accretion disk is set at $10^{19}$~cm$^{-3}$, the iron abundance at 1.0 and the ionization parameter at $\log\xi = 3.5$.
It is clear that the \textsc{xillverns} model exhibits a flatter reflection spectrum. 
The flat shape of the \textsc{xillverns} spectrum makes the reflection component prominent or dominating at low energies, which could lead to the misfitting of the accretion disk emission. 
The electron free-free radiation is an efficient mechanism to dissipate energy (cooling) in the case of the thin disk~\citep{1973A&A....24..337S, 2010ApJ...718..695G}, resulting in the generation of more low-energy photons that flatten the reflection spectrum in the model \textsc{xillverns}~\citep{2022ApJ...926...13G}.
However, during the super-Eddington phase, the radiation pressure could suppress electromagnetic collisions between electrons and charged particles. In this scenario, the electron free-free radiation ceases to be an efficient mechanism for dissipating energy. 
This suggests that the contribution of electron free-free radiation should be reconsidered when the luminosity is high.

\subsection{The geometry of accretion disk change during the state transition}
\label{sec:The geometry of accretion disk}

We find a clear state transition from super-Eddington to sub-Eddington accretion during the outburst decay phase. 
As shown in Figure~\ref{fig:fig5_relxillns}, \ref{fig:fig5_reflionx}, and especially \ref{fig:fig_index_incl}, the emissivity index of the disk shows a clear difference before and after the state transition.
Previous studies find that the emissivity profile of the disk is affected by the height of the X-ray illuminator above the disk plane~\citep{2012MNRAS.424.1284W, 2013MNRAS.430.1694D}.
As the illuminator height decreases, the emissivity profile becomes steeper.
This effect also appears in the neutron star accretion disks~\citep{2018MNRAS.475..748W}.
During the super-Eddington period, the inner region of the disk becomes thick due to the radiation pressure. This leads to height differences between different radii regions of the inner disk and makes the emissivity profile flat. 
As the source progresses into the sub-Eddington period, the accretion disk gradually becomes thinner, resulting in minimal height differences across different regions of the disk. This leads to a significant increase in the emissivity index~\citep{2012MNRAS.424.1284W, 2013MNRAS.430.1694D}.
The large emissivity index ($> 6.0$) during the sub-Eddington period is similar to that in a neutron-star system when a hotspot or belt close to the equator illuminates the disk and the equator is very close to the inner edge of the disk \citep{2018MNRAS.475..748W}.
Hence, the transition from a supercritical, thick accretion disk to a thinner one provides a natural explanation for the observed evolution of the emissivity index.

\section{Conclusions}

In this work we have carried out a detailed analysis of the black hole X-ray transient 4U~1543$-$47 as observed during its outburst in 2021 with \textit{Insight}-HXMT. Our conclusions are:

\begin{enumerate}[1.]
    \item We find a clear state transition during the outburst decay in the hardness-intensity diagram of the source (Figure~\ref{fig:fig_hid_fit}). Considering the black hole mass and distance, 9.4 $M_\odot$ and 7.5 kpc, respectively, when the source crosses this transition the luminosity is close to  the Eddington limit. 
    \item The 2--10 keV light curves before and after the state transition can be fitted by two exponential functions with a short ($\sim 16$ days) and a long ($\sim 130$ days) decay time scales, respectively (Figure~\ref{fig:fig_hid_fit}). The long decay time scale in the soft state indicates that a large amount of accretted material is stored during the quiescence state for $\sim 20$ years.
    \item  
    A soft excess or warm corona, which can be characterized by Comptonization with a low electron temperature of approximately 2.0 keV, is observed exclusively during the super-Eddington state (Figure~\ref{fig:thcomp_frac}), similar to the “hypersoft” state of the 2005 super-Eddington outburst of the BHB GRO~J1655$-$40.
    We suggest that this soft excess could be produced by the accretion disk itself in the inner radiation-pressure dominated region of the supercritical disk (section~\ref{sec:The super-Eddington state}).
    \item Strong reflection features are detected in all observations. 
    The reflection can be fitted with the model \textsc{relxillns} or \textsc{reflionx\_bb}. \textsc{relxillns} tends to overestimate the flux at low energies (section~\ref{sec:The super-Eddington state}).  
    \item 
    The best-fitting emissivity index of the disk also shows a clear difference before and after the transition, from $\sim 3.0-5.0$ in the super-Eddington period to $\sim 7.0-9.0$ in the sub-Eddington period (Figure~\ref{fig:fig_index_incl}). Based on the returning disk radiation, the change of the thickness of the accretion disk could explain the change of the emissivity index (section~\ref{sec:The geometry of accretion disk}).  
    \item Our results during the state transition indicate that the accretion disk may have changed significantly near the Eddington limit. We propose that the transition is primarily driven by a change from a supercritical to a thin disk.

\end{enumerate}

\section*{Acknowledgements}

We thank the Reviewer for helpful comments, both of science and language.
We thank prof. Lian Tao, Shujie Zhao for help in all aspects.
We thank Menglei Zhou for helpful discussion.
This work made use of data from the \textit{Insight}-HXMT mission, a project funded by China National Space Administration (CNSA) and the Chinese Academy of Sciences (CAS).
This work is supported by the National Key R\&D Program of China (2021YFA0718500). 
GZ acknowledges funding support from the National Natural Science Foundation of China (NSFC) under grant No. U1838116. 
GB acknowledges the science research grants from the China Manned Space Project.
YZ acknowledges support from the China Scholarship Council (CSC 201906100030), and the Dutch Research Council (NWO) Rubicon Fellowship, file no.\ 019.231EN.021. 
MM acknowledges the research programme Athena with project num-ber 184.034.002, which is (partly) financed by the Dutch Research Council (NWO). 
This material is based upon work supported by Tamkeen under the NYU Abu Dhabi Research Institute grant CASS.

\section*{Data Availability}

The data analyzed in this paper are available at the official website of \textit{Insight}-HXMT~\url{http://HXMTweb.ihep.ac.cn}.



\bibliographystyle{mnras}
\bibliography{4U_1543} 




\appendix

\section{systematic errors}
\label{sec:systematic errors}

In our spectral analysis part, we add systematic errors (1\% for LE, 2\% for ME, provided by the \textit{Insight}-HXMT instrument team) to the \textit{Insight}-HXMT spectra. Taking sub-exposure P030402600120 as an example, we get the $\chi^2/{\rm d.o.f.}$ $=$ 81/213 by fitting with the \textsc{relxillns} model. The reduced $\chi^2$ is $\sim 0.5$, significantly less than 1, suggesting that the systematic errors provided by the HXMT instrument team are overestimated.

\begin{figure}
	\includegraphics[width=\columnwidth]{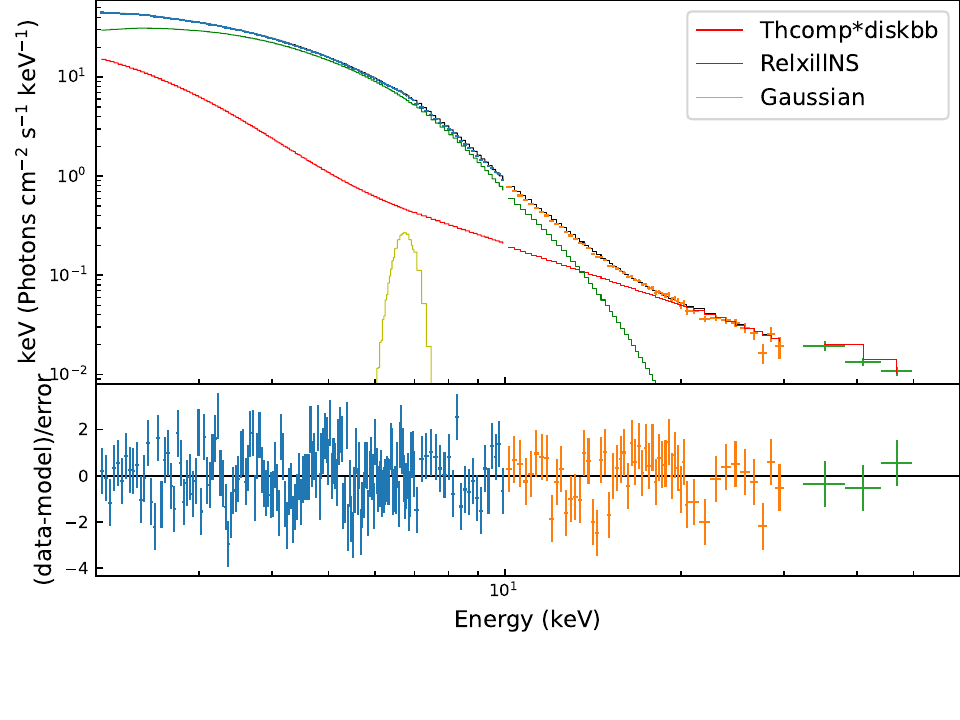}
	\vspace{-12mm}
    \caption{The unfolded spectrum and the residuals of sub-exposure P030402600120 (MJD 59382) without LE/ME systematic error. The model is \textsc{constant*tbabs*(thcomp*diskbb+relxillns)}. The contributions to the model of the various additive components are also plotted.}
    \label{fig:without LE/ME systematic}
\end{figure}

To explore the effect of the systematic errors we fit the spectra of sub-exposure P030402600120 without adding LE/ME systematic error to the data and using the same model. We get a $\chi^2/{\rm d.o.f.} = 268/213$. The best-fitting model and residuals are shown in Figure~\ref{fig:without LE/ME systematic}. The spread of the residuals is around 0 and random. The residuals are similar to that of the bottom panel of Figure~\ref{fig:fig4}, but the fluctuation of the residuals of \textit{Insight}-HXMT/LE in the 2--5 keV range are larger without adding systematic error.
A comparison is shown in Table \ref{tab:table-P030402600120} of the effect of systematic errors on the spectral parameters. We can see that the parameters are consistent regardless of whether the systematic errors are added or not.

\begin{figure}
	\includegraphics[width=\columnwidth]{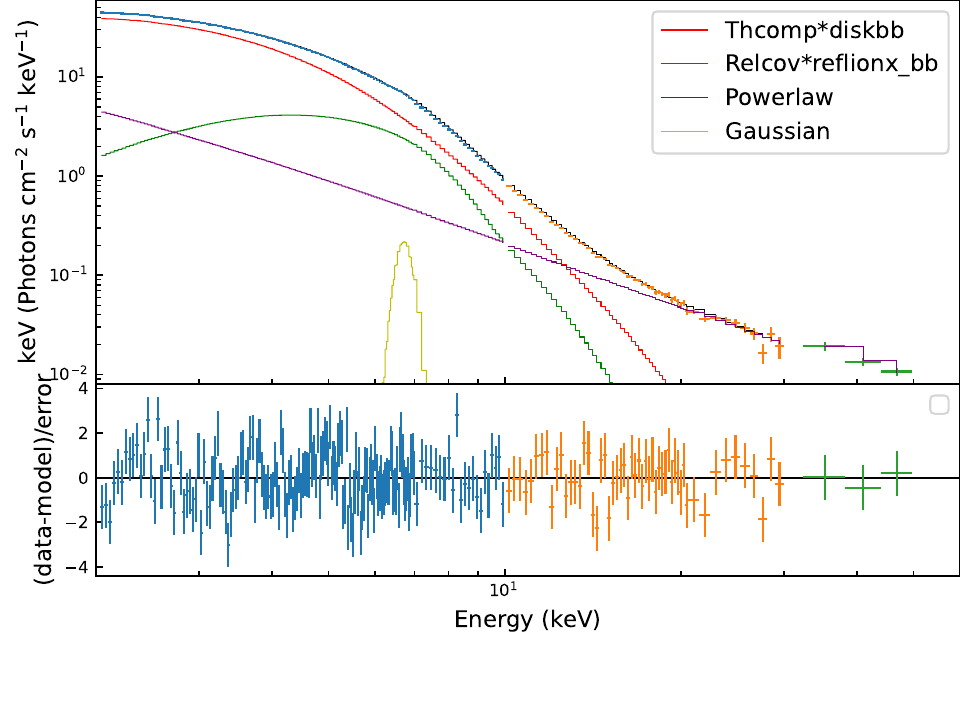}
	\vspace{-12mm}
    \caption{The unfolded spectrum and the residuals of sub-exposure P030402600120 (MJD 59382) without LE/ME systematic error. The model is \textsc{constant*tbabs*zxipcf*(thcomp*diskbb+powerlaw+} \textsc{relconv*reflionx\_bb+gaussian)}. The contributions to the model of the various additive components are also plotted.}
    \label{fig:without LE/ME systematic_reflionxbb}
\end{figure}

The same processing is also applied to \textsc{reflionx\_bb}. Without applying systematic error for the LE/ME spectra, the $\chi^2/{\rm d.o.f.}$ is 264/214. The best-fitting model and residuals are shown in Figure~\ref{fig:without LE/ME systematic_reflionxbb}
and the spectral parameters are recorded in Table~\ref{tab:table-P030402600120_reflionxbb}.

\begin{table}
	\centering
	\caption{The spectral parameters of sub-exposure P030402600120. The model is \textsc{constant*tbabs*(thcomp*diskbb+relxillns+gaussian)}.}
	\label{tab:table-P030402600120}
\renewcommand\arraystretch{1.5}
\begin{tabular}{ccc}
\hline 
Parameter & With systematic errors & Without systematic errors \\ 
\hline 
\multicolumn{3}{c}{\textsc{thcomp*diskbb}}\\
$\Gamma$ & $2.8_{-0.1}^{+0.1}$ & $2.8_{-0.2}^{+0.1}$\\ 
$kT_{\rm e}$ (keV) & $>58$ & $>15$\\
cov\_frac & $0.18_{-0.04}^{+0.03}$ & $0.18_{-0.08}^{+0.10}$\\
$T_{\rm in}$ (keV) & $0.55_{-0.04}^{+0.03}$ & $0.55_{-0.01}^{+0.01}$\\
Norm ($\times 10^3$) & $63.8_{-13.6}^{+18.7}$ & $63.4_{-6.1}^{+5.3}$\\
\hline 
\multicolumn{3}{c}{\textsc{relxillns}} \\
Index & $3.5_{-0.1}^{+0.3}$ & $3.5_{-0.2}^{+0.2}$\\ 
Inclination angle ($^{\circ}$) & 35(fixed) & 35(fixed)\\
$kT_{\rm bb}$ (keV) & $1.15_{-0.01}^{+0.01}$ & $1.14_{-0.01}^{+0.01}$\\
$\log\xi$ & $3.57_{-0.05}^{+0.02}$ & $3.57_{-0.02}^{+0.02}$\\
$\log N$ (cm$^{-3}$) & $>18.8$ & $>18.9$\\
Norm & $0.27_{-0.01}^{+0.01}$ & $0.27_{-0.01}^{+0.01}$\\
\hline
\multicolumn{3}{c}{\textsc{guassian}}\\
Line energy (keV) & 6.72$_{-0.05}^{+0.05}$ & 6.71$_{-0.03}^{+0.03}$\\
Sigma & 0.28$_{-0.07}^{+0.08}$& 0.26$_{-0.05}^{+0.05}$\\
Norm & 0.028$^{+0.006}_{-0.007}$ & 0.027$^{+0.004}_{-0.004}$\\
\hline
\end{tabular}
\end{table}

\begin{table}
	\centering
	\caption{The spectral parameters of sub-exposure P030402600120. The model is \textsc{constant*tbabs*zxipcf*(thcomp*diskbb+powerlaw+}  \textsc{relconv*reflionx\_bb+gaussian)}.}
	\label{tab:table-P030402600120_reflionxbb}
\renewcommand\arraystretch{1.2}
\begin{tabular}{ccc}
\hline 
Parameter & With systematic errors & Without systematic errors \\ 
\hline
\multicolumn{3}{c}{\textsc{zxipcf}}\\
nH & 6.7$^{+3.8}_{-2.2}$ & 4.8$^{+0.9}_{-0.8}$\\
$\log\xi$ & 1.0(fixed) & 1.0(fixed) \\
cov\_fraction &  0.18$^{+0.10}_{-0.05}$ & 0.29$^{+0.09}_{-0.07}$\\
\hline
\multicolumn{3}{c}{\textsc{thcomp*diskbb}}\\
$\Gamma$ & 2.6$^{+0.8}_{-0.5}$ & 3.1$^{+0.7}_{-0.5}$\\ 
$kT_{\rm e}$ (keV) & 2.0(fixed) & 2.0(fixed) \\
cov\_frac & 0.08$^{+0.12}_{-0.04}$ & 0.14$^{+0.11}_{-0.7}$\\
$T_{\rm in}$ (keV) & 1.00$^{+0.02}_{-0.01}$ & 0.97$^{+0.03}_{-0.03}$\\
Norm ($\times 10^3$) &  16.3$^{+1.9}_{-1.0}$ & 18.9$^{+2.5}_{-2.4}$\\
\hline 
\multicolumn{3}{c}{\textsc{powerlaw}}\\
$\Gamma$ & 2.8$^{+0.4}_{-0.6}$  & 3.1$^{+0.2}_{-0.1}$\\
Norm  & 7.9$^{+39.8}_{-7.0}$ & 29.1$^{+33}_{-14}$ \\
\hline
\multicolumn{3}{c}{\textsc{relconv*reflionx\_bb}} \\
Index & 3.0$^{+0.2}_{-0.3}$ & 3.1$^{+0.1}_{-0.1}$\\ 
Inclination angle ($^{\circ}$) & 30(fixed) & 30(fixed)\\
$kT_{\rm bb}$ (keV) & \multicolumn{2}{c}{link to $T_{\rm in}$}\\
$\log\xi$ & $>3.3$ & $>3.9$\\
$\log N$ (cm$^{-3}$) & 20(fixed) & 20(fixed)\\
Norm & 5.1$^{+1.1}_{-1.8}$ & 6.0$^{+0.6}_{-0.6}$\\
\hline
\multicolumn{3}{c}{\textsc{guassian}}\\
Line energy (keV) & 6.7(fixed)& 6.7(fixed)\\
Sigma & 0.2(fixed)& 0.2(fixed)\\
Norm & 0.019$^{+0.004}_{-0.007}$ & 0.017$^{+0.002}_{-0.002}$\\
\hline
\end{tabular}
\end{table}

\section{Parameter errors obtained with MCMC}

\begin{figure*}
	\includegraphics[width=1.09\textwidth]{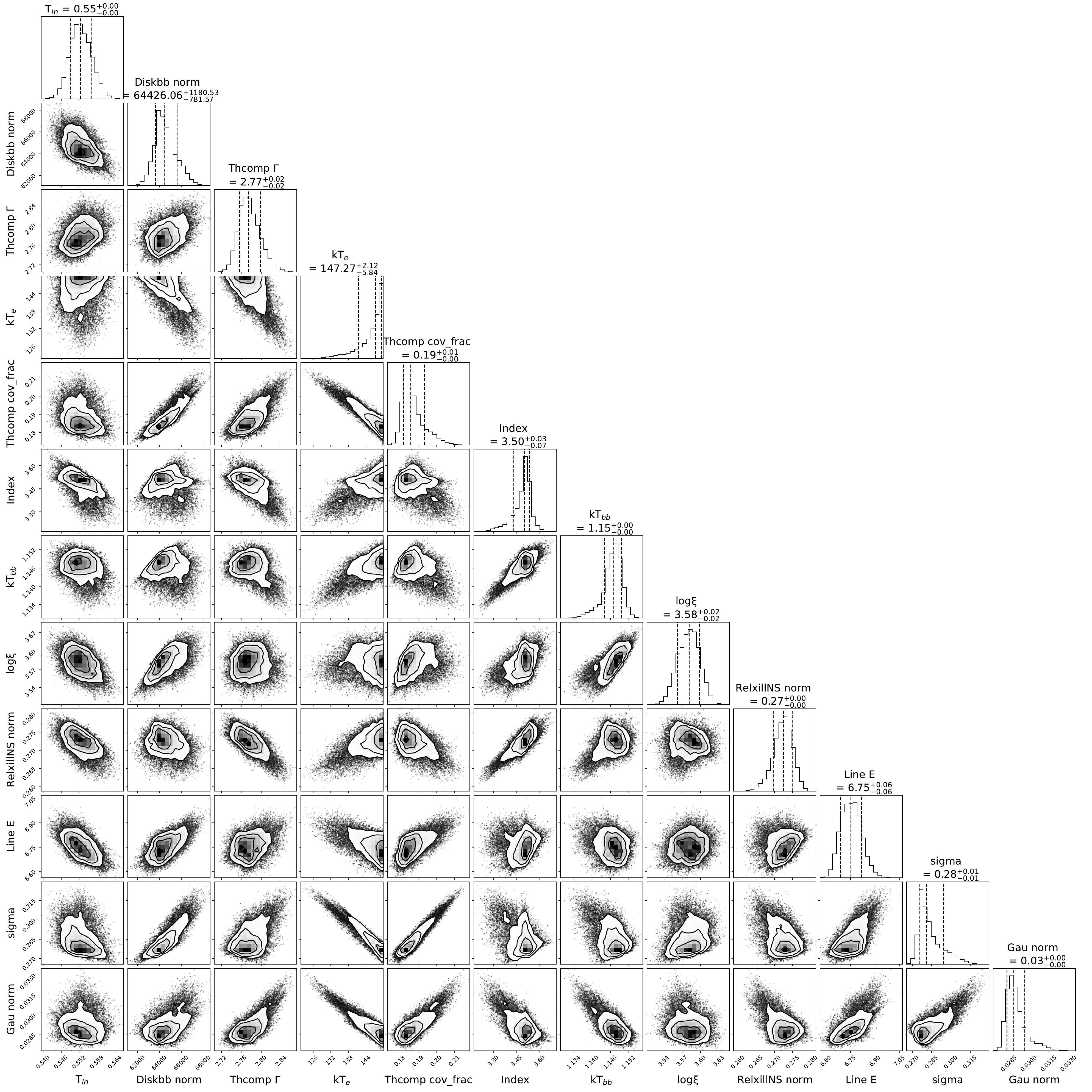}
    \caption{The probability distribution obtained by MCMC of the spectral parameters of sub-exposure P030402600120 (MJD 59382). The model is \textsc{constant*tbabs*(thcomp*diskbb+relxillns+gaussian)}.The density of the accretion disk always pegs to the upper limit so that the chain does not give its distribution.}
    \label{fig:mcmc_relxillns}
\end{figure*}

\begin{figure*}
	\includegraphics[width=1.09\textwidth]{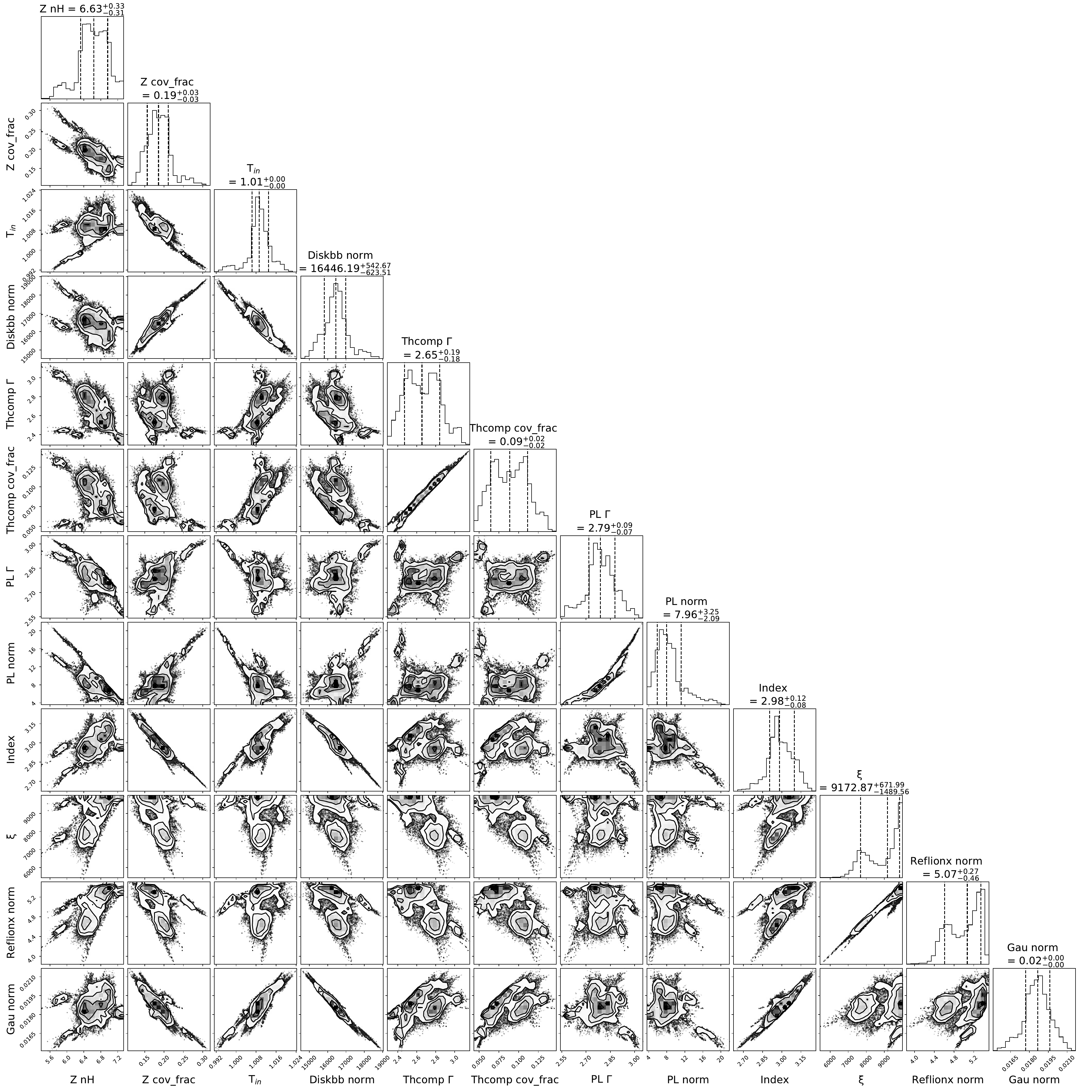}
    \caption{The probability distribution obtained by MCMC of spectral parameters of sub-exposure P030402600120 (MJD 59382). The model is \textsc{constant*tbabs*zxipcf*(thcomp*diskbb+powerlaw+} \textsc{relconv*reflionx\_bb+gaussian)}.}
    \label{fig:mcmc_reflionx}
\end{figure*}

We also calculate the spectral parameter errors using the Markov chain Monte-Carlo algorithm (MCMC). The Goodman-Weare algorithm with 8 walkers is applied for a total of 100000 samples.
We set a burn-in phase of 50000, so that the chain can reach a steady state.
We present the results obtained from sub-exposure P030402600120 using both reflection models in Figure~\ref{fig:mcmc_relxillns} with \textsc{relxillns} and Figure~\ref{fig:mcmc_reflionx} with \textsc{reflionx\_bb}, respectively. The error ranges for the spectral parameters at the 1-sigma confidence level are shown above the probability distribution maps. 
Noticeably, the uncertainty ranges of the parameters obtained by minimum Chi-Square are wider than those obtained through MCMC.
Even though some parameters, especially the emissivity index that is critical for assessing the disk geometry, appear to be highly generated with others, the error bars are not really large. For instance the emissivity index still within approximately 3-4.

\section{The spectral parameters}
\label{sec:The spectral parameters}

We show the spectral parameters of the first four sub-exposures with \textit{Insight}-HXMT of 4U 1543$-$47 and give an online link to all observations, in Table~\ref{tab:the spectral parameters} with \textsc{relxillns} and Table~\ref{tab:the spectral parameters_reflionx} with \textsc{reflionx\_bb}, respectively.

\begin{table*}
\caption{The spectral parameters of observations with \textit{Insight}-HXMT of 4U 1543$-$47. The model is \textsc{constant*tbabs*(thcomp*diskbb+} \textsc{relxillns+gaussian)}.}
     \label{tab:the spectral parameters}
\setlength{\tabcolsep}{1.35mm}{     
\renewcommand\arraystretch{1.5}
\begin{tabular}{ccccccccccccc} 
\hline 
Time & exp-ID & $T_{in}$ & \textsc{diskbb} & $\Gamma$ & cov\_frac & Index & $T_{bb}$ & $log\xi$ & \textsc{relillns} & lineE & \textsc{gaussian} & $\chi^2$ \\ 
 (MJD) &  & (keV) & norm ($10^3$) &  &  &  & (keV) &  & norm & (keV) & norm ($10^{-1}$) & \\
\hline 
59379.7 & P030402600101 & $0.71\pm0.06$ & $28_{-8}^{+6}$ & $4.73_{-0.19}^{+0.33}$ & $>0.82$ & $3.5\pm0.3$ & $1.22\pm0.02$ & $3.47\pm0.07$ & $0.265\pm0.019$ & $6.74\pm0.08$ & $0.132_{-0.048}^{+0.047}$ &0.53 \\ 
\hline 
\multirow{1}{*}{59379.9} & \multirow{1}{*}{P030402600102} & $0.61\pm0.09$ & $52_{-21}^{+30}$ & \multirow{1}{*}{$4.67_{-0.30}^{+0.07}$} & \multirow{1}{*}{$>0.61$} & \multirow{1}{*}{$2.9\pm0.1$} & $1.17\pm0.02$ & \multirow{1}{*}{$3.61\pm0.08$} & $0.259\pm0.016$ & \multirow{1}{*}{$6.77\pm0.07$} & $0.161_{-0.045}^{+0.044}$ & \multirow{1}{*}{0.46} \\ 
\hline 
\multirow{1}{*}{59380.0} & \multirow{1}{*}{P030402600103} & $0.66\pm0.08$ & $35_{-10}^{+7}$ & \multirow{1}{*}{$4.37_{-0.13}^{+0.23}$} & \multirow{1}{*}{$>0.72$} & \multirow{1}{*}{$3.3\pm0.3$} & $1.19\pm0.02$ & \multirow{1}{*}{$3.48\pm0.07$} & $0.271\pm0.013$ & \multirow{1}{*}{$6.83\pm0.06$} & $0.204_{-0.054}^{+0.043}$ &\multirow{1}{*}{0.59} \\ 
\hline 
\multirow{1}{*}{59380.2} & \multirow{1}{*}{P030402600104} & $0.69\pm0.08$ & $30_{-9}^{+10}$ & \multirow{1}{*}{$4.49_{-0.36}^{+0.26}$} & \multirow{1}{*}{$>0.50$} & \multirow{1}{*}{$3.6\pm0.3$} & $1.21\pm0.02$ & \multirow{1}{*}{$3.45\pm0.07$} & $0.272\pm0.015$ & \multirow{1}{*}{$6.77\pm0.07$} & $0.210_{-0.054}^{+0.052}$ &\multirow{1}{*}{0.54} \\ 
\hline 
\multicolumn{13}{c}{more information see \textsc{the spectral parameters with relxillns.txt}} \\
\hline 
\end{tabular}}
\end{table*}

\begin{table*}
\caption{The spectral parameters of observations with \textit{Insight}-HXMT of 4U 1543$-$47. The model is \textsc{constant*tbabs*zxipcf*(thcomp*diskbb+} \textsc{powerlaw+relconv*reflionx\_bb+gaussian)}.}
     \label{tab:the spectral parameters_reflionx}
\setlength{\tabcolsep}{0.6mm}{     
\renewcommand\arraystretch{1.5}
\begin{tabular}{cccccccccccccc} 
\hline 
Time & exp-ID & \textsc{zxipcf} & \textsc{zxipcf} & Tin & \textsc{diskbb} & \textsc{thcomp} & \textsc{powerlaw} & \textsc{powerlaw}  & Index & $\xi$ & \textsc{relfionx\_bb} & \textsc{gaussian} & $\chi^2$ \\ 

(MJD) &  & nH & cov\_frac & (keV) & norm ($10^3$) & cov\_frac & photon index & norm &  & ($10^2$) & norm & norm ($10^{-1}$)  &  \\ 
\hline 
59379.7 & P030402600101 & $6.04_{-0.52}^{+1.03}$ & $0.19\pm0.02$ & $1.00\pm0.01$ & $19.0_{-0.1}^{+0.7}$ & $0.111\pm0.004$ & 2.5 &$<0.076$ & $2.8\pm0.1$ & $>95.8$ & $6.43_{-0.13}^{+0.11}$ & $0.059_{-0.038}^{+0.038}$ & 0.53 \\ 
\hline 
59379.9 & P030402600102 & $7.15_{-0.64}^{+0.75}$ & $0.20$ & $1.00\pm0.01$ & $19.7_{-0.1}^{+0.2}$ & $0.098\pm0.003$ & 2.5 &$<0.020$ & $2.6\pm0.1$ & $>98.5$ & $5.90_{-0.27}^{+0.12}$ & $0.058_{-0.043}^{+0.044}$ & 0.52 \\ 
\hline 
59380.0 & P030402600103 & $5.99_{-1.81}^{+1.46}$ & $0.21\pm0.08$ & $0.99\pm0.05$ & $19.8_{-2.3}^{+3.1}$ & $0.124\pm0.009$ & 2.5 &$<0.201$ & $2.9\pm0.3$ & $72.7_{-23.8}^{+27.3}$ & $6.15_{-0.77}^{+1.88}$ & $0.108_{-0.061}^{+0.070}$ & 0.61 \\ 
\hline 
59380.2 & P030402600104 & $5.57_{-1.82}^{+1.51}$ & $0.24\pm0.9$ & $0.96\pm0.04$ & $22.1_{-4.6}^{+6.1}$ & $0.119\pm0.009$ & 2.5 &$0.103_{-0.037}^{+0.043}$ & $3.0\pm0.2$ & $>65.5$ & $7.70_{-1.37}^{+0.79}$ & $0.132_{-0.064}^{+0.042}$ & 0.48 \\ 
\hline 
\multicolumn{14}{c}{more information see \textsc{the spectral parameters with reflionx\_bb.txt}} \\
\hline 
\end{tabular}}
\end{table*}


\bsp	
\label{lastpage}
\end{document}